\documentclass[11pt]{article}
\usepackage{epsfig}
\usepackage{amsmath,amssymb}
\usepackage{mathtools}
\usepackage{braket}
\usepackage{hyperref}
\usepackage{authblk}
\usepackage{graphicx}
\usepackage{comment}
\usepackage{xspace}
\usepackage{xcolor}
\usepackage{float}
\usepackage{wrapfig}
\let\OldS\S
\renewcommand{\S}{\OldS\xspace}
\graphicspath{ {./images/} }
\usepackage[section]{placeins}
\usepackage{color}
\DeclareMathOperator{\csch}{csch}

\begin{document}
\date{}
\title{Entanglement inequalities for timelike intervals within dynamical holography}

\author[1]{Gaurav Katoch\thanks{gauravitation@gmail.com}}
\author[1]{ Debajyoti Sarkar\thanks{dsarkar@iiti.ac.in}}
\author[1]{Bhim Sen\thanks{bhimsen10496@gmail.com}}

\affil[1]{Department of Physics\\

Indian Institute of Technology Indore\\

 Khandwa Road, Indore, 453552, India}

\maketitle

\begin{abstract}
This paper extends our previous work (arXiv:2504.14313) of a single timelike subregion to two, in the framework of AdS$_3$-Vaidya holography. We confirm the positivity of timelike mutual information and the statement of weak monotonicity when the subregions are non-overlapping. We also study entanglement inequalities such as Araki-Lieb inequality and strong subadditivity when the intervals start to overlap. In line with the recent findings in the literature, we provide explicit working examples showing that the timelike version of the strong subadditivity is generally violated in these setups, even though the statements of subadditivity and Araki-Lieb inequality hold true.
\end{abstract}

\vspace{10mm}
Keywords: AdS$_3$-Vaidya, Timelike mutual information, Strong subadditivity.

\newpage
\tableofcontents 
\newpage

\section{Introduction}\label{sec:intro} 

It won't be an overstatement to say that perhaps the most attractive feature of holography \cite{Maldacena:1997re, Gubser:1998bc, Witten:1998qj} is that it provides the geometrization of field theory quantities, and thereby supplies an alternative tractable handle on complicated field theoretic relationships governing them. One prime example of this is the geometrization of entanglement entropy (EE) of the boundary field theory in  terms of the bulk Ryu-Takayanagi (RT) surfaces \cite{Ryu:2006bv}. This quantity has been studied in exhaustive detail, and many of its features and distinctions are entailed via strong subadditivity, Araki-Lieb inequality, weak monotonicity, monogamy of mutual information etc. These properties are already of utmost significance purely within the subject of quantum information theory \cite{Witten:2018zva}, but they can also be rigorously worked out staying within the holographic setup \cite{Headrick:2007km,Headrick:2013zda}. To appreciate the simplicity and geometric intuition that the holography offers, for comparison, one can be invited to have a look at the highly non-trivial proof of strong subadditivity contained in  \cite{lieb1973fundamental}. \\

Despite having such desirable features, holographic entanglement entropy is riddled with short-distance divergences proportional to the area of the RT surface, and any regulator dependent methods to render it finite essentially ends up making it scheme dependent. 
However, one can unequivocally define a scheme independent linear combination of entanglement entropies (called the mutual information) for two disjoint systems, that can be rendered free of UV divergences due to cancellation of short-distance singularities. Mutual information provides a measure of degree of correlation only between the disjoint subsystems and not with the rest of the system. It roughly tries to answer the question of how much information does the two disjoint systems share.   Moreover, mutual information $I(A,B)$ can be used to study the evolution of correlations in non-equilibrium systems. Other than being UV finite, it also satisfies a positivity statement ($I(A,B)\geq 0$), with the inequality saturating to zero only when the density matrix of the combined interval is factorizable. In other words, mutual information vanishes only when the two systems lose their mutual correlation. This latter setup is often called a `disconnected' phase, where the combined EE of the two intervals is simply a sum of their individual EE with the rest of the system. A positive mutual information therefore occurs in the `connected' phase.\\

All of these above notions can be straightforwardly generalized to the case when the intervals are timelike. Therefore, if one instead considers two non-overlapping timelike subregions $A$ and $B$ anchored at different boundary times, one can investigate the behaviour of timelike version of mutual information defined to be  
 \begin{align}
  \tilde{I}&\equiv   \tilde{S}(A)+\tilde{S}(B)-\tilde{S}(A\cup B)\,.\label{TMI}
 \end{align}
 Above, the various $\tilde{S}$ denote the real part of the corresponding timelike entanglement entropy (TEE).\footnote{TEE is a complex quantity. For time independent states, in equations like \eqref{TMI}, the imaginary parts cancel out. However, as we will see, this is not generally the case for the dynamical scenarios. In most of our analyses in this paper, we will always strip away the imaginary parts from the various TEEs of interest and just focus on their real parts.} The discussion of such timelike mutual information (TMI) made its first appearance in \cite{Doi:2023zaf}, where it was defined by replacing the usual entanglement entropies (in mutual information) by  their timelike counterparts.   The above expression for TMI can also be written alternatively in a more suggestive manner, namely
  \begin{align}
    \tilde{I}&\equiv   \tilde{S}^D_{AB}-\tilde{S}(A\cup B)\,.
 \end{align}
Here $\tilde{S}^D_{AB}=\tilde{S}(A)+\tilde{S}(B)$ denotes the collective timelike entanglement entropy in the disconnected phase, and $\tilde{S}(A\cup B)$ is the collective TEE of the dominating  phase (the phase with lesser entropy). The dominating phase can either be the disconnected phase or the connected phase depending upon the separation between the two subsystems. Following the work of \cite{Doi:2023zaf}, one can show that for static cases (particularly for boundaries dual to pure AdS or BTZ black hole \cite{Banados:1992wn}), TMI behaves exactly like its spacelike counterpart. In particular, when the two timelike intervals are separated by large times, the dominating phase is the disconnected one, yielding a vanishing TMI. However as their temporal separation decreases, TMI hits a turning point at which the connected phase becomes the dominating one, rendering a monotonically increasing positive TMI.\\

In a previous paper \cite{Katoch:2025bnh}, following the works of \cite{Doi:2022iyj,Doi:2023zaf} in static states, we computed the holographic timelike entanglement entropy (HTEE) of a globally quenched CFT dual to AdS$_3$-Vaidya spacetime resulting from the spherical null shell collapse.  
Our holographic computations closely parallelled the covariant generalization (but now for the timelike case) of the RT prescription (known as HRT surfaces \cite{Hubeny:2007xt}) that one implements for the spacelike intervals. In the original works of \cite{Doi:2022iyj,Doi:2023zaf} the TEE was inspired from the `pseudo entropy' developed in the context of non-hermitian states, which also appears in the discussions of de Sitter/conformal field theory (dS/CFT) correspondence \cite{Narayan:2015vda,Narayan:2017xca,Doi:2022iyj,Doi:2023zaf}. However, there is no formal proof (unlike e.g. the proofs of \cite{Lewkowycz:2013nqa,Dong:2016hjy} for the spacelike case) of the holographic TEE prescription for the dynamical cases. Therefore, it would be important to subject the validity of our results (particularly for the dynamical case) to test by investigating whether they also obey the analogous quantum information theoretic inequalities as mentioned above. \\

For example, for the spatial subregions, the validity of strong subadditivity (SSA) in the static spacetimes was proven holographically in \cite{Headrick:2007km} using the dual RT surfaces (with a more detailed analysis presented in \cite{Headrick:2013zda}).\footnote{Quite interestingly, an alternative approach to various holographic entropy inequalities has been proposed using the deterministic method based on the existence of a contraction map  in \cite{Bao:2024azn,Bao:2025sjn}. } The question of its validity for time dependent spacetimes like AdS$_3$-Vaidya was carried out in \cite{Callan:2012ip,Grimaldi:2025jad} using the relevant HRT prescriptions.\footnote{Purely within quantum information theory, the statement of strong subadditivity is closely related to the concavity of von Neumann entropy, and in a way uniquely characterizes the von Neumann entropy by providing the strongest constraining condition (together with the continuity of the eigenvalues of density matrix, and its invariance under unitary transformations \cite{Hirata:2006jx}).} However, the status of SSA for pseudo entropy is still in a relatively shaky ground. Although it lacks ample examples, it was shown in \cite{Mollabashi:2021xsd} (for free massive scalar fields living on the asymptotic boundary of Lifshitz spacetime), that pseudo entropy may sometimes violate the SSA constraint. 

It is therefore natural to expect that holographic TEE (being a form of pseudo entropy), will also violate SSA on general grounds (see also \cite{Jiang:2025pen,Afrasiar:2025eam} for similar statements). 

In our present work, we indeed find explicit evidence for this statement of violation of SSA, while still satisfying the validity of other `weaker' constraints, such as positivity of TMI and the timelike version of Araki-Lieb inequality. In turn, this also puts our earlier results \cite{Katoch:2025bnh} on a firmer ground.  \\

\textbf{Plan of the paper:} The plan of this paper is as follows. We start in section \ref{sec:Recap} with a brief summary of our findings in \cite{Katoch:2025bnh}, which considered a single timelike interval in the AdS$_3$-Vaidya background. The later parts of the paper will rely heavily on the results mentioned here, when we consider an additional interval and study their combined entanglement structure. In section \ref{sec: TMI}, we will then proceed with the studies of timelike mutual information as we introduce the second, disjoint interval in the picture. It immediately gives rise to a number of distinct configurations, that depend upon whether the combined system is in a connected or in the disconnected phase. We will find that in all the configurations, the timelike mutual information behaves as in the spacelike case, in that it gets positive and monotonically increasing once it crosses from disconnected to the connected phase. In section \ref{sec: SSA}, particularly in subsections \ref{ssat}  and \ref{subsec:SSAlarget}, we will start with the investigation of strong subadditivity inequalities, by making the intervals overlap by a fixed amount. Enumerating all the possible configurations, we find that whereas the subadditivity is valid throughout, the SSA may undergo a violation. We provide more evidence on the violation of SSA in subsection \ref{sec:SSAVtchange}, where we vary the overlap as well. Finally, we conclude in section \ref{sec:conclude}. Appendix \ref{a} provides some useful formulas that we have used in section \ref{sec: SSA}.

\section{Review: single timelike interval in AdS$_{3}$-Vaidya}\label{sec:Recap}

For the convenience of the reader, in this section we have put together a summary of our trailing work \cite{Katoch:2025bnh} that would be necessary throughout the paper.  
The metric that we employ is the AdS$_{2+1}$-Vaidya spacetime 
   \begin{align}
        ds^2=\frac{L^2}{z^2}\left(-f(v,z)\,dv^2-2dvdz+dx^2\right)\,.
    \end{align}
The Vaidya metric describes a null shell composed of tensionless dust. The blackening function is given by $f(v,z)=1-M(v)z^2$, where we have
    \begin{align}
          \begin{array}{cc}
 M(v)\equiv M \Theta(v)=\Bigg\{ & 
\begin{array}{cc}
 M \qquad &\text{for}\qquad v>0 \\
 0 \qquad &\text{for}\qquad  v<0
\\
\end{array}
 \\
\end{array}\nonumber
    \end{align}
  
The infalling matter is concentrated with a delta function spike at $v=0$, because we are interested in the zero thickness limit of the collapsing null shell.\footnote{It takes a certain amount of time for the pure AdS state to settle down to BTZ after a quench. It is a standard result of sudden approximation of time dependent perturbation theory, see e.g. Chapter $XVII$, \OldS 8 of \cite{messiah1999quantum}. We will neglect this technicality throughout our analysis.} Here, we have defined the ingoing Eddington-Finkelstein coordinate $v$ as 
         \begin{align}
        v\equiv T-\int^z_0 \frac{dz'}{f(z')}\nonumber\,.
        \end{align}
Henceforth, in this and the following analysis, we have set the scale by plugging AdS radius $L$ to unity. The horizon is located at $z_h=\frac{1}{\sqrt{M}}$  and the inverse temperature of the ($2+1$)-dimensional black hole is given by $\beta=\frac{2\pi}{\sqrt{M}}$. This metric is an uplift from the AdS$_3$ Poincar\'{e} patch (with coordinates $(T,z,x)$) to a dynamical one, due to the perturbation by $M$. \\

In our previous work, we  evaluated the quantum extremal surfaces corresponding to timelike subregions anchored at different times at the boundary. Using their equations, we also computed the timelike entanglement entropy for the dynamically evolving geometry dual to the relaxing boundary theory. As the interval is timelike itself, it passes through various `cases', as it moves up in the boundary time. Our main objective was to capture the evolution of this holographic TEE for these various cases.  In this set up, the geometry at the earliest times starts off as pure AdS (labelled case 1), and at the late times relaxes approximately to the eternal BTZ (labelled case 4). labelling the lower and the upper endpoints of the time interval by $T_i$ and $T_j$ respectively, we have the following four cases:
\\

\textbf{Case 1:} When the interval is purely in the AdS region ($T_i<T_j<0$), the resulting HTEE is given by 
\begin{align}
    \tilde{S}_{1}(T_i,T_j)&=\frac{c}{3}\log\left(\frac{T_j-T_i}{\epsilon}\right)+\frac{c}{6}\pi i\,. \label{case1}
\end{align}
This is nothing but the result of \cite{Doi:2022iyj,Doi:2023zaf}, which was carried out for the static case ($c$ being the central charge of the CFT$_2$, and $\epsilon$ is a UV regulator). As we will now see, during the course of its evolution to eventual relaxation (case 4), the HTEE  trajectory goes through two other interpolating cases (labelled by case 2 and case 3 in our pervious work as well) as discussed below.
\\

\textbf{Case 2:} It arises when the boundary subregion is straddling the null shock in such a manner that more than half of the interval is in the AdS region i.e. $T_j<|T_i|$. In other words, the mid-point of the subregion interval is located below the collapsing null shell.  This case is characterized by the upper spacelike extremal surface crossing over to the AdS region through null shell at $z=z_c$, and leads to a dynamical evolution that exhibits the growth phase of TEE before passing on to the next case. The TEE for this case is given by 
\begin{align}
    \tilde{S}_{2}(T_i,T_j)=\frac{c}{6}\log \left(\frac{\beta }{\pi \epsilon }\sinh \left(\frac{\pi}{\beta}\left(T_j-T_i\right)\right)\right)+\frac{c}{6}\log\left(\frac{T_j-T_i}{ \epsilon}\right)+\frac{c}{6}
\Phi_2(T_i,T_j)+\frac{c}{6}\pi i\,.\label{case2}
\end{align}\\
Here we have repackaged the term responsible for the dynamical evolution  as 
  {\small{  \begin{align}
   &\Phi_2(T_i,T_j)=-\sinh ^{-1}\left(\frac{1}{2 \sqrt{\sinh ^2\left(\frac{\sqrt{M} T_i}{2}\right) \sinh ^2\left(\frac{\sqrt{M} T_j}{2}\right) \text{csch}^2\left(\frac{1}{2} \sqrt{M} \left(T_i-T_j\right)\right) \text{csch}^2\left(\frac{1}{2} \sqrt{M} \left(T_i+T_j\right)\right)}}\right)\nonumber\\
    &+\sinh ^{-1}\Bigg[\frac{1}{4} \sqrt{M} \sinh \left(\frac{1}{2} \sqrt{M} \left(T_i+T_j\right)\right) \csch\left(\frac{\sqrt{M} T_i}{2}\right)\csch\left(\frac{\sqrt{M} T_j}{2}\right)\times \nonumber\\ &\sqrt{\frac{\left(T_i+T_j\right) \left(\sqrt{M} \left(T_i+T_j\right)-4 \coth \left(\frac{1}{2} \sqrt{M} \left(T_i+T_j\right)\right)+4 \cosh \left(\frac{1}{2} \sqrt{M} \left(T_i-T_j\right)\right) \csch\left(\frac{1}{2} \sqrt{M} \left(T_i+T_j\right)\right)\right)}{\sqrt{M}}}\Bigg]\,.
\end{align}}}
 This term arises from the limits of integration being the crossing point $z_c$ on the null shell. This expression is responsible for governing the growth phase of TEE at the end of case 1. \\

\textbf{Case 3:} As the boundary interval evolves to larger times, eventually the upper extremal surface crosses the null shell and falls into the singularity region. This new case takes over the moment $T_j>|T_i|$, or in other words, when the mid-point of the interval crosses and goes above the null shell.  The expression for holographic TEE for this case 3  is 
\begin{align}
    \tilde{S}_{3}(T_i,T_j)=\frac{c}{6}\log \left(\frac{\beta }{\pi \epsilon }\sinh \left(\frac{\pi}{\beta}\left(T_j-T_i\right)\right)\right)+\frac{c}{6}\log \left(\frac{T_j-T_i}{\epsilon}\right)+\frac{c}{4}i\pi\,. \label{case3}
\end{align}
Due to the absence of shell-crossing, this behaviour is static in nature and displays a plateau region in the evolution of TEE.  \\

\textbf{Case 4:} The final case 4 kicks in when the subregion is fully anchored on the BTZ region of the dynamical geometry and is characterized by $T_j>T_i>0$. The resulting expression for the HTEE is 
\begin{align}
  \tilde{S}_{4}(T_i,T_j)&= \frac{c}{3} \log \left(\frac{\beta }{\pi \epsilon }\sinh \left(\frac{\pi}{\beta}\left(T_j-T_i\right)\right)\right) +\frac{c}{6}\Phi_4(T_i,T_j)+\frac{c\pi i}{4}\label{case4}\,,
\end{align} 
with 
{\small{\begin{align}
&\Phi_4(T_i,T_j)=-\text{csch}^{-1}\left(2 \sinh \left(\frac{\sqrt{M} T_i}{2}\right) \sinh \left(\frac{\sqrt{M} T_j}{2}\right) \text{csch}\left(\frac{1}{2} \sqrt{M} \left(T_j-T_i\right)\right) \text{csch}\left(\frac{1}{2} \sqrt{M} \left(T_i+T_j\right)\right)\right)\nonumber\\
&-\sinh ^{-1}\Bigg[\frac{1}{4} \sqrt{M} \sinh \left(\frac{1}{2} \sqrt{M} \left(T_i+T_j\right)\right) \text{csch}\left(\frac{\sqrt{M} T_i}{2}\right) \text{csch}\left(\frac{\sqrt{M} T_j}{2}\right) \times\nonumber\\
&\sqrt{\frac{\left(T_i+T_j\right) \left(\sqrt{M} \left(T_i+T_j\right)-4 \coth \left(\frac{1}{2} \sqrt{M} \left(T_i+T_j\right)\right)+4 \cosh \left(\frac{1}{2} \sqrt{M} \left(T_j-T_i\right)\right) \text{csch}\left(\frac{1}{2} \sqrt{M} \left(T_i+T_j\right)\right)\right)}{\sqrt{M}}}\Bigg]\nonumber\,.
\end{align}}}

\begin{figure}[h]
    \centering
    \includegraphics[width=0.6\linewidth]{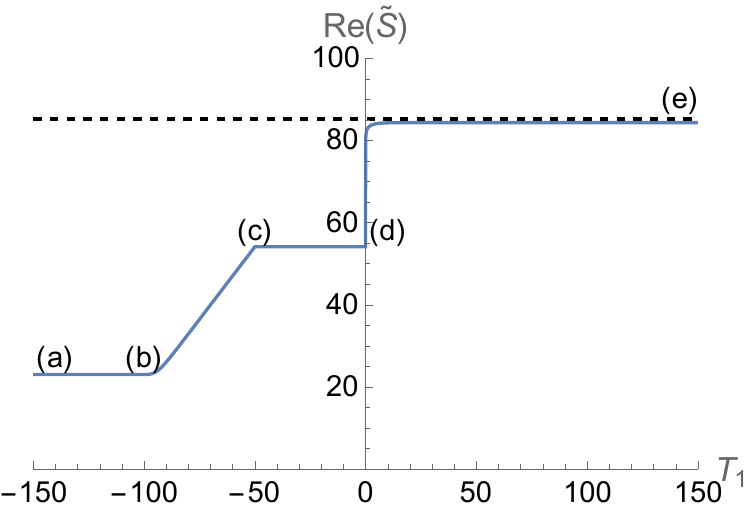}
    \caption{Evolution of TEE as the function of boundary time. The different cases correspond to the segments (a)-(b), (b)-(c), (c)-(d) and (d)-(e) respectively. We have plotted it for the subregion size $T_2-T_1=100$, $M=0.5$ and $\epsilon=0.001$.}
    \label{fig:TEE}
\end{figure}

Curiously, even though in this case the lower spacelike branch crosses the shell, it immediately asymptotes to the black hole horizon (and stays effectively right outside the horizon). In the HTEE, this is reflected by a steep jump from case 3 (provided by the function $\Phi_4(T_j,T_i)$), leading to the saturation of HTEE at the late times. The saturation value is essentially the expected HTEE  of the BTZ geometry \cite{Doi:2022iyj,Doi:2023zaf}. A combined plot of all these separate evolutions can be seen in figure \ref{fig:TEE}. 

\section{Distinct configurations for timelike mutual information}\label{sec: TMI}

In this section, we will concentrate on two disjoint identical boundary timelike intervals and the resulting time-dependent effects encoded in mutual information. Because we will never consider more than two intervals, it is a good idea to clarify our notations at this point. For calculational simplicity, we will always consider two intervals $A=(T_1,T_2)$ and $B=(T_3,T_4)$ to be of the same size $\tau$.\footnote{As it will be clear later, the total number of configurations drastically proliferates if this is not the case. Although this adds in more intricate details, the point we want to make can be achieved by considering identical intervals.} The separation between the endpoints $T_2$ and $T_3$ will be denoted by $t$. The intervals will always be ordered according to $T_4>T_3>T_2>T_1$ with $T_2=T_1+\tau,$ $T_3=T_2+t=T_1+\tau+t$ and $T_4=T_3+\tau=T_1+2\tau+t$. 
\\

In fact, it is quite straight-forward to have an estimate of the behaviour of TMI for the relatively simpler static cases of empty AdS and BTZ. In both the cases, it can be easily shown that as the distance between the intervals gets smaller, we switch from the disconnected ($D$) to the connected ($C$) phase, echoing what happens for the spacelike case \cite{Doi:2023zaf}.\footnote{Because for these above static cases, the TEE can be understood as coming from a Wick rotation from the spacelike case \cite{Doi:2022iyj,Doi:2023zaf,Das:2023yyl}, this result is almost automatic.} For example, from \eqref{case1} one can easily see that the disjoint sum $\tilde{S}^{D}_{AdS}=\tilde{S}(A)+\tilde{S}(B)$ leads to
  \begin{align}
     \tilde{S}^{D}_{AdS}&=\frac{2c}{3}\log\left(\frac{\tau}{\epsilon}\right)+\frac{c}{3}\pi i\,.
 \end{align}
 Whereas if it were in the connected phase, the resulting TEE will be given by
  \begin{align}
     \tilde{S}^{C}_{AdS}&=\frac{c}{3}\log\left(\frac{2\tau+t}{\epsilon}\right)+\frac{c}{3}\log\left(\frac{t}{\epsilon}\right)+\frac{c}{3}\pi i\,.
 \end{align}
Therefore because of the monotonicity of the logarithmic dependence upon the arguments, it is manifest from the expressions above that as $t$ approaches smaller and smaller values, we will have $\tilde{S}^{C}_{AdS}\leq \tilde{S}^{D}_{AdS}$ . Hence, in the disconnected phase when $\tilde{S}(A\cup B)=\tilde{S}^D_{AB}$, TMI identically vanishes due to \eqref{TMI}. Whereas, in the connected phase, $\tilde{S}(A\cup B)=\tilde{S}^C_{AB}$ and therefore, $\tilde{I}^C=\tilde{S}^D_{AB}-\tilde{S}^C_{AB}\geq 0$. This connected phase value of the TMI can be simply calculated, and is given by
\begin{align}\label{iads}
 \tilde{I}^C_{AdS}&=\frac{c}{3}\log\left(\frac{\tau^2}{t(2\tau+t)}\right)\,.
\end{align}
This is positive definite as $t\to 0$. \\
  
A similar computation in the eternal BTZ then recovers the results of \cite{Doi:2023zaf}. In particular, for large temporal separations, the disconnected phase dominates yielding a vanishing TMI. But as the intervals are brought nearer, the system transitions into a connected phase with 
\begin{align}
     \tilde{I}^C_{BTZ}=\frac{c}{3}\log\left(\frac{\sinh \left(\frac{\pi}{\beta}\tau\right)\sinh \left(\frac{\pi}{\beta}\tau\right)}{\sinh \left(\frac{\pi}{\beta}(2\tau+t)\right)\sinh \left(\frac{\pi}{\beta}t\right)}\right)\,.
\end{align}
Once again, this is positive definite as $t\to 0$. Interestingly, in both of these static cases the imaginary parts cancel out explicitly. As we will see shortly, for the Vaidya spacetime, the above two scenarios only appear asymptotically, with a multitude of different configurations appearing during the intermediate times. In some of these cases the imaginary parts do not cancel out. However, as mentioned in the introduction, for all the cases we will only be interested in the real parts of the HTEE. 
\\

\begin{figure}[H]
    \centering
    \includegraphics[width=1\linewidth]{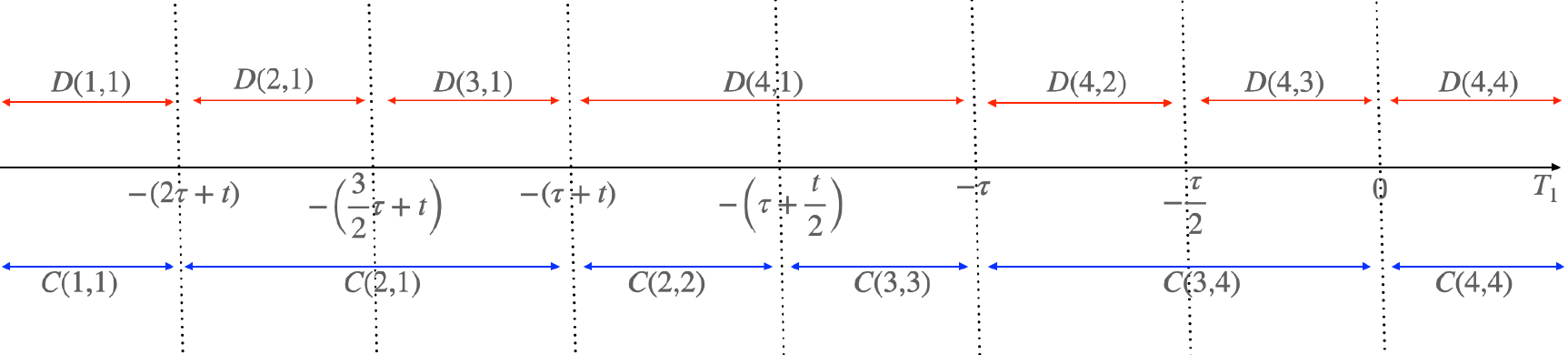}
    \caption{Timeline of various disconnected and connected phases with respect to $T_1$.}
    \label{fig:MItimeline}
\end{figure}

It turns out that the real dynamical evolution of TMI for the late time  BTZ formation is quite complex and far more interesting than the static case. Chronologically proceeding, there arises as far as six other intermediate configurations interpolating between the pure AdS in the far past to the  BTZ in the late future. We have listed each of those configurations in the figure \ref{fig:MItimeline} (not to scale) along with their corresponding ranges of validity with respect to the boundary time $T_1$. 

A few more words on our notations are in order. In above, we adopted the labelling system in which  $D$ and $C$ denote the disconnected and the connected phases respectively. On the other hand, the values of $(n,m)$ gives us the case numbers of the relevant interval in question. The disconnected phase is particularly simple to understand. Here $n$ corresponds to the case that the upper interval $B$ is in, and $m$ for the case of the interval $A$ (case numbers are what appear in section \ref{sec:Recap}). For example, in the configuration $D(n,m)$, the resulting HTEE will be given by 
\begin{align}
    \tilde{S}_{D(n,m)}=\tilde{S}_{n}(T_3,T_4)+\tilde{S}_{m}(T_1,T_2)\,.\label{D}
\end{align}
In order to compute $\tilde{S}_n(T_3,T_4)$, we'll simply make the substitution of $j=4, i=3$ (for example, for $n=3$, it will be the real part of \eqref{case3}). On the other hand, $\tilde{S}_m(T_1,T_2)$ can be computed by substituting $j=2,i=1$ for any $m=1,2,3,4$ (using the equations \eqref{case1} to \eqref{case4} as appropriate). \\

The $C(n,m)$ phase is slightly different, as here the resulting intervals are $(T_1,T_4)$ and $(T_2,T_3)$, with $(n,m)$ denoting their case numbers respectively. Therefore, in order to write the required HTEE, we will use 
\begin{align}
    \tilde{S}_{C(n,m)}=\tilde{S}_{n}(T_1,T_4)+\tilde{S}_{m}(T_2,T_3)\,.\label{C}
\end{align}
For example, $\tilde{S}_2(T_1,T_4)$ will be given by \eqref{case2} with the substitution $j=4$ and $i=1$. Similarly, any other entropy values can be obtained from using equations \eqref{case1} to \eqref{case4}.
 Throughout the section, without compromising precision, we will rely on the numerical  approach and strip away the imaginary part to adopt the statement \eqref{TMI} for the real part of TMI while ascertaining the existence of finite $t_c> 0$. While plotting we have consistently used subsystem size $\tau=100$, $M=0.5$  and $\epsilon=10^{-3}$ in the units of AdS curvature.
\\

\subsection{Earliest time Configuration  $(D(1,1)-C(1,1))$  }

\begin{figure}[H]
    \centering
    \includegraphics[width=0.3\linewidth]{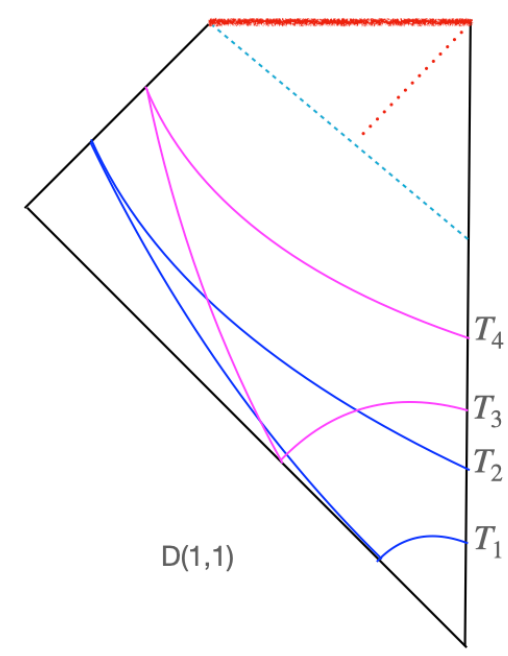}
\hspace{3cm}
\includegraphics[width=0.3\linewidth]{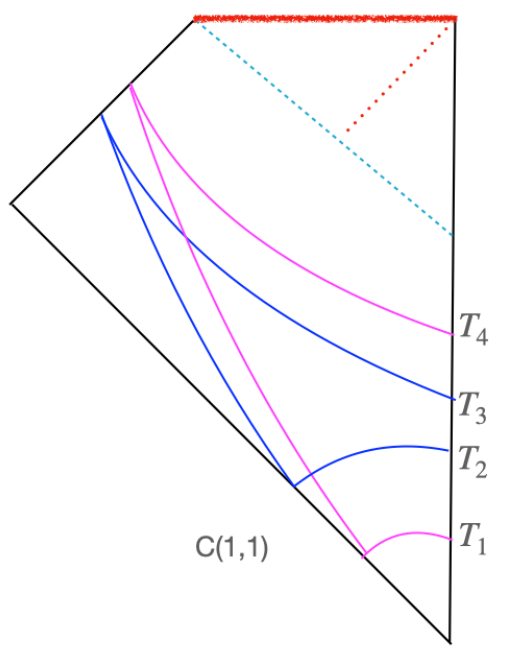}{}\caption{  $D(1,1) - C(1,1)$ geodesic configurations valid through $-\infty < T_1<-(2\tau+t)$.}
    \label{fig:1d1c}
\end{figure}

The very first configuration of geodesics that arises, pertains to the situation when the intervals corresponding to the subregions are located in the far past of the collapsing shell. Effectively, the spacetime is pure AdS as the collapse of the null shell to BTZ has not yet taken place. Therefore, the geodesics emanating from the respective end points of the timelike subregions are restricted to lie entirely in the AdS spacetime.
This first configuration  is characterized by the endpoint of the upper subregion $B$ to be located at $T_4=T_1+2\tau+t<0$, or in terms of $T_1$
$$T_1<-(2\tau+t)\,.$$
When the subsystem $B$ and subsystem $A$ are separated by a large enough distance $t$, the disconnected phase is the dominant phase. The configuration of geodesics is shown in the left panel of the adjoining figure \ref{fig:1d1c} and pertains to each subsystem being individually in case 1. As discussed earlier, the disconnected configuration is therefore denoted by $D(1,1)$.  In this case, the total TEE for both the subsystems is given by \eqref{D} for $n=m=1$ and is
 \begin{align}
        \tilde{S}_{1D}
         &=\frac{c}{3}\log\left(\frac{\tau^2}{\epsilon^2}\right)+\frac{c}{3}\pi i\,.
         \end{align}

         The corresponding connected phase develops as the separation between the subsystems is made smaller by fixing $T_1$ and decreasing $T_4$. The connected configuration is still $C$(1,1)  and is shown in the right panel of figure \ref{fig:1d1c}. Total TEE of the system can be governed from  \eqref{C} after replacing $n=m=1$  to give  
         \begin{align}
         \tilde{S}_{1
         C}
         &=\frac{c}{3}\log\left(\frac{ (2\tau +t)t}{\epsilon^2}\right)+\frac{c}{3}\pi i.
    \end{align}
 
\begin{figure}[H]
            \centering       \includegraphics[width=0.5\linewidth]{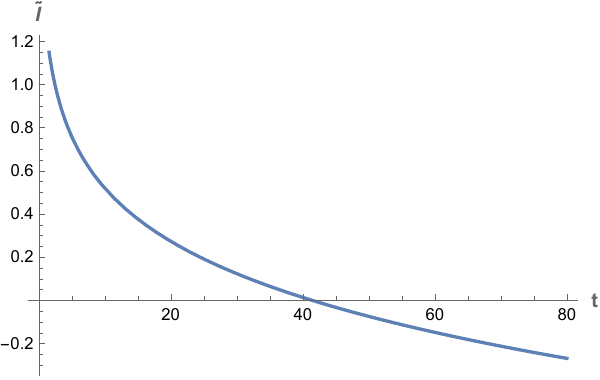}
            \caption{$C(1,1)\to D(1,1)$ phase transition in TMI occurring at $t_c=41.4$ for $T_1=-250$. This is precisely the value one obtains by equating the mutual information \eqref{iads} to zero, which is $t_c=(\sqrt{2}-1)\tau.$}
            \label{fig:MI1}
        \end{figure}

The critical subsystem separation $t_c$ for which the phase transition takes place can simply be figured out by equating the value of TEE for disconnected phase to the connected phase, i.e. the moment TMI equates to zero. For the case at  hand, due to the simplicity of the mathematical expressions it is possible to solve for $t_c$. However in the latter cases, we will adopt numerical approach to pinpoint the critical separation from the plots of the TMI. For the present case, TMI follows the behaviour given in figure \ref{fig:MI1} which shows the phase transition from the connected phase to the disconnected phase as the separation is made to grow larger. 

\subsection{Second Configuration ($D(2,1) - C(2,1)$)   }

\begin{figure}
    \centering
    \includegraphics[width=0.3\linewidth]{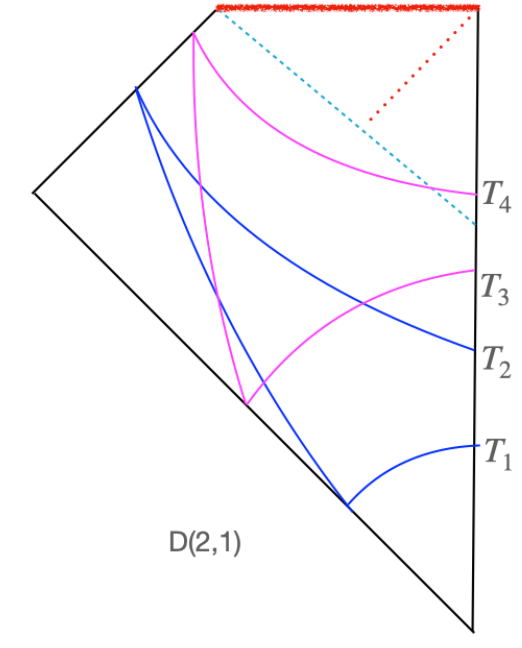}
\hspace{3cm}
\includegraphics[width=0.3\linewidth]{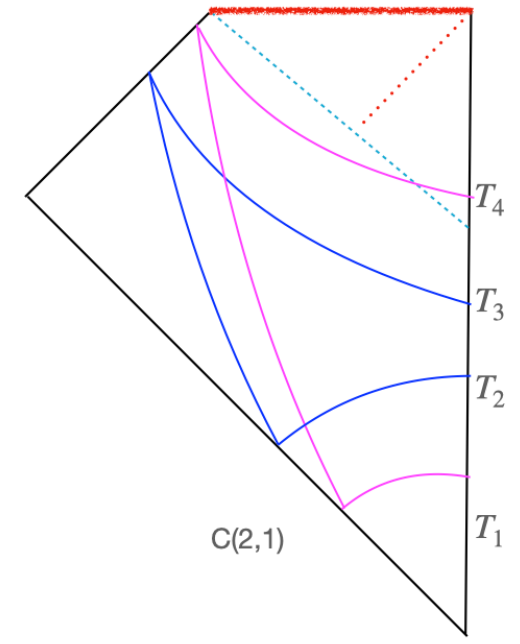}{}\caption{$D(2,1)-C(2,1)$ geodesic configurations valid through $-(2\tau+t)<   T_1<-\left(\frac{3\tau}{2}+t\right)$.}
    \label{fig:2d2c}
\end{figure}

As the boundary subsystems are translated along the boundary time, the second configuration arises, in which the upper subsystem crosses the shell and its upper end point gets anchored at the value $T_4>0$ in a manner such that $T_4<|T_3|$. While the lower subsystem still has access to only the AdS side of the geometry. To get the range of validity of this configuration,  we notice that second configuration starts off the moment configuration $D(1,1)-C(1,1)$ ends, i.e. at $T_1>-(2\tau+t)$. Afterwards, it sustains until the mid-point of the transitioning subregion is lying just below the null shell. 
In other words, we demand that $\frac{T_4+T_3}{2}<0$ giving us the condition on $T_1<-\left(\frac{3\tau}{2}+t\right)$. 
It is therefore clear that the subregions $B$ and $A$ are in cases 2 and 1 respectively, and therefore it will be denoted by $D(2,1)$ when we consider the disconnected phase. In summary, the configuration $D(2,1)-C(2,1)$ is valid through the following time period
\begin{align*}
 -(2\tau+t)<   T_1<-\left(\frac{3\tau}{2}+t\right)\,.
\end{align*}
A little thought also shows that when in connected phase, the subregion $(T_1,T_4)$ will be in case 2 and the interval $(T_2,T_3)$ will be in case 1. Therefore, the connected phase will be denoted by $C(2,1)$. We have provided a schematic diagram of both of these disconnected and connected setups in figure \ref{fig:2d2c}.

In this case, the total TEE in the disconnected phase can be computed from \eqref{D} for $n=2$ and $m=1$, and is given by
    \begin{align}
        \tilde{S}_{2D}
&=\frac{c}{6}\log \left(\frac{\beta }{\pi \epsilon }\sinh \left(\frac{\pi}{\beta}\tau\right)\right)+\frac{c}{2}\log\left(\frac{\tau}{ \epsilon}\right)+\frac{c}{6}
\Phi_2(T_3,T_4)+\frac{c}{3}\pi i\,.
    \end{align}
Similarly, the total TEE for the connected phase $C(2,1)$ can be worked out simply from \eqref{C} by making appropriate replacements of $n=2$ and $m=1$ to give 
    \begin{align}
        \tilde{S}_{2C}
&=\frac{c}{6}\log \left(\frac{\beta }{\pi \epsilon }\sinh \left(\frac{\pi}{\beta}\left( 2\tau+t\right)\right)\right)+\frac{c}{6}\log\left(\frac{ (2\tau+t) t^2}{ \epsilon^3}\right)+\frac{c}{6}
\Phi_2(T_1,T_4)+\frac{c}{3}\pi i\,.\label{2C}
    \end{align}
    
\begin{figure}[h!tbp]
        \centering        \includegraphics[width=0.5\linewidth]{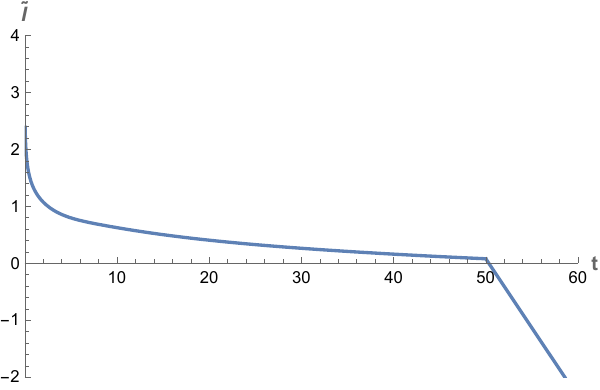}
        \caption{$C(2,1)\to D(2,1)$ phase transition in TMI occurring at $t_c=50.3$ for $T_1=-200.01$.   }
        \label{fig:MI2}
    \end{figure}

As the interval between the subsystems is varied, the phase transition takes place from the connected to the disconnected phase when the separation grows to $t_c$. The value of $t_c$ can be read off from the adjoining plot in \eqref{fig:MI2}, where the timelike mutual information vanishes. \\

We must remark here that as the set of intervals are evolved further in the boundary time, the connected and the disconnected configurations  operate independently of each other. For example, as we evolve the system further, although the disconnected phase $D(2,1)$  switches to the $D(3,1)$, but the connected phase still remains in $C(2,1)$. 

\subsection{Third Configuration $(D(3,1) - C(2,1))$ }
\begin{figure}[H]
    \centering    \includegraphics[width=0.3\linewidth]{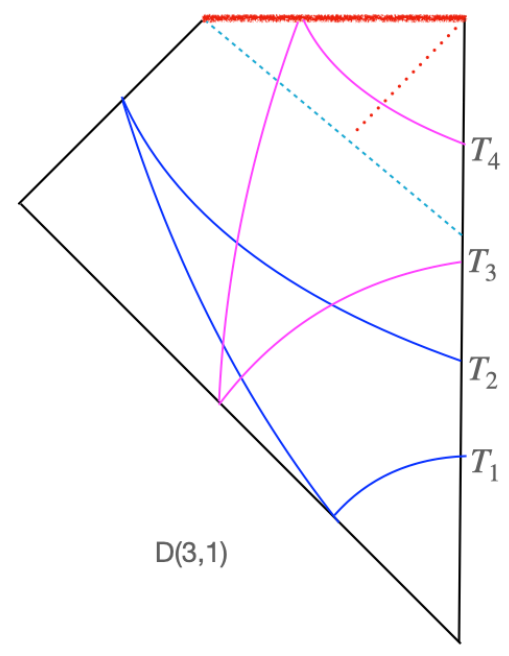}
\hspace{3cm}
\includegraphics[width=0.3\linewidth]{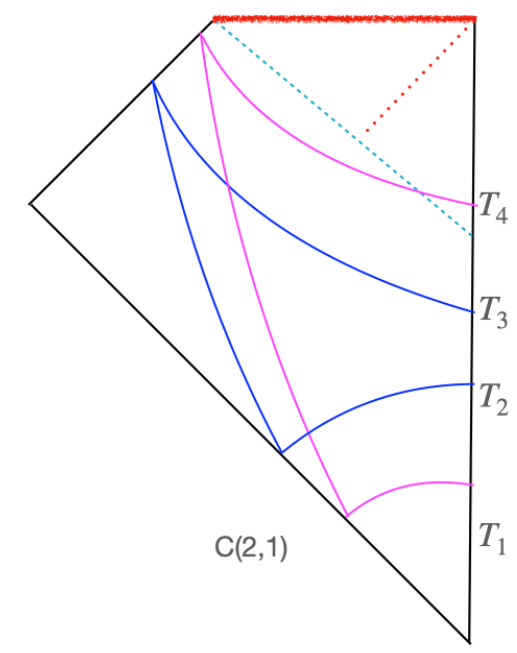}{}\caption{$D(3,1)-C(2,1)$ geodesic configurations valid through $-\left(\frac{3\tau}{2}+t\right) <T_1<-(\tau+t)$.}
    \label{fig:3d3c}
\end{figure}

As is clear from figure \ref{fig:3d3c}, the next  configuration appears when the lower subregion $A$ is still in the AdS part of the geometry, while the upper subregion $B$ has transitioned into case 3. Therefore, the corresponding  disconnected phase should appropriately be labelled by $D(3,1)$. This phase is characterised by the fact that the mid-point of the straddling interval $B$ is above the null shell and has now become positive, while the lower end point $T_3$ is still below the shell. In other words, the range of validity of configuration $D(3,1)$ is 
\begin{align*}
    -\left(\frac{3\tau}{2}+t\right) <T_1<-(\tau+t)\,.
\end{align*}

The combined TEE for the disconnected phase can be computed from \eqref{D} for $(n,m)=(3,1)$ and is given by
   \begin{align}
          \tilde{S}_{3D}
          &=\frac{c}{6}\log \left(\frac{\beta }{\pi \epsilon }\sinh \left(\frac{\pi}{\beta}\tau\right)\right)+\frac{c}{2}\log \left(\frac{ \tau}{\epsilon}\right)+\frac{5c}{12}\pi i\,.
    \end{align}
   
    \begin{figure}[h!tbp]
        \centering
        \includegraphics[width=0.5\linewidth]{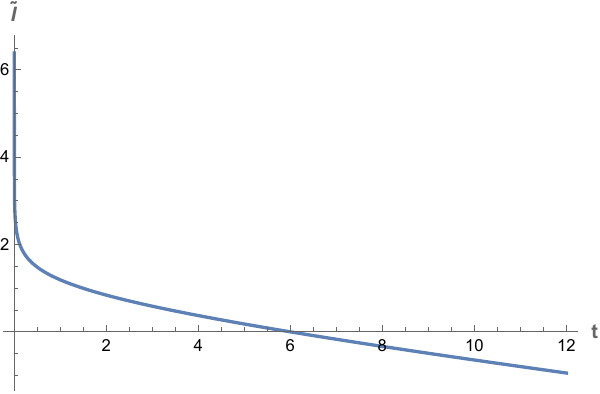}
        \caption{Phase transition $C(2,1)\to D(3,1)$ at critical separation $t_c=5.9$ for $T_1=-149.01$.  }
        \label{fig:MI3}
    \end{figure}
Given the ranges of time intervals, it is clear that when considering the connected phase, the corresponding configuration is still the trailing one $C$(2,1). Hence, the total TEE for the connected phase is still given by \eqref{2C}. The resulting plot for the mutual information has been provided in figure \ref{fig:MI3}.
From the plot, we see that as the intervals are dragged apart, the connected phase transitions to the disconnected phase at a finite value  of $t_c$.
\\

\subsection{Fourth Configuration $(D(4,1) - C(2,2)$ or $C(3,3))$ }

 Next follows the fourth disconnected configuration $D$(4,1), in which the upper subregion $B$ fully transitions into the BTZ side of the geometry, while the lower subregion $A$ is still completely in the AdS part of the geometry.  It is characterized by 
 \begin{align}
     -(\tau+t)<T_1<-\tau\,.\nonumber
 \end{align}
  \begin{figure}[H]
        \centering
        \includegraphics[width=0.3\linewidth]{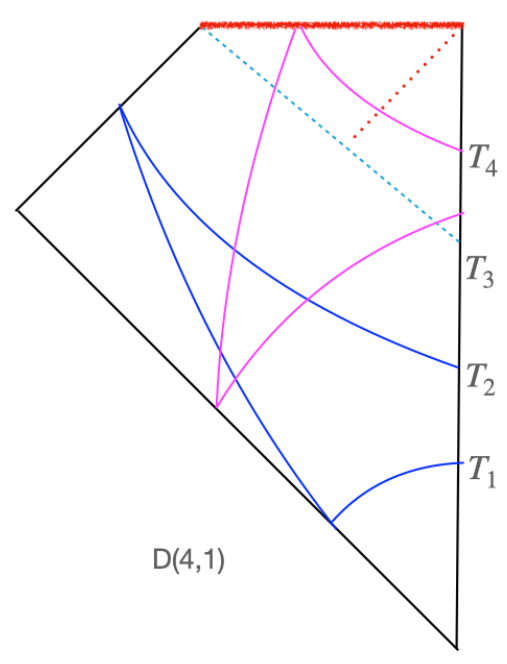}
       \hfill\includegraphics[width=0.3\linewidth]{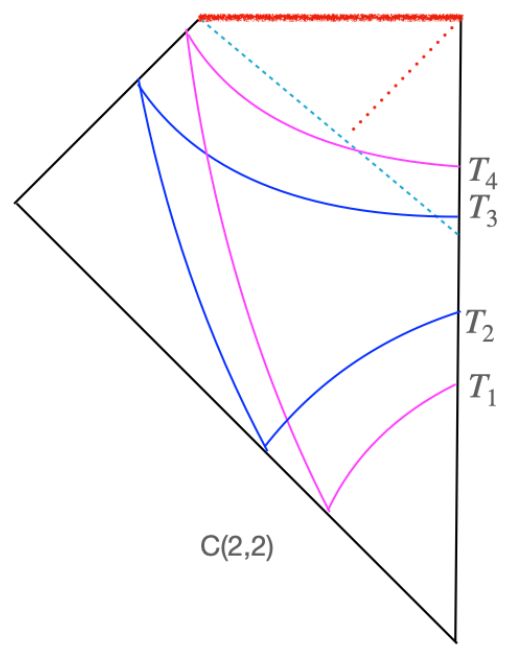}
       \includegraphics[width=0.3\linewidth]{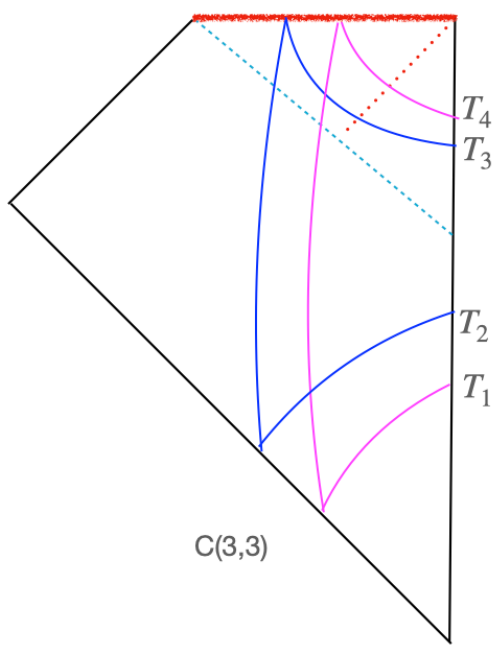}
        \caption{$D(4,1)-C$(2,2) (left and middle panels), $D(4,1)-C(3,3)$ (left and right panels) geodesic configurations valid through $ -(\tau+t)<T_1<-(\tau+\frac{t}{2})$ and $-\left(\tau+\frac{t}{2}\right)<T_1<-\tau$ respectively.}
        \label{fig:4d4c5c}
    \end{figure}
 The total TEE of the corresponding disconnected phase is given by plugging $(n,m)=(4,1)$ in \eqref{D} to give
\begin{align}
        \tilde{S}_{4D}
        &= \frac{c}{3} \log \left(\frac{\beta }{\pi \epsilon }\sinh \left(\frac{\pi}{\beta} \tau \right)\right)+\frac{c}{3}\log\left(\frac{ \tau}{\epsilon}\right) +\frac{c}{6}\Phi_4(T_3,T_4,\beta)+\frac{5c}{12}\pi i\,.
    \end{align}
   
\begin{figure}[H]
    \centering
    \includegraphics[width=0.39\linewidth]{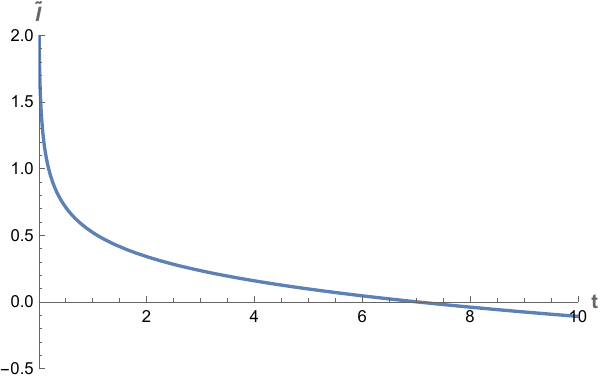}
\hspace{3cm}
\includegraphics[width=0.39\linewidth]{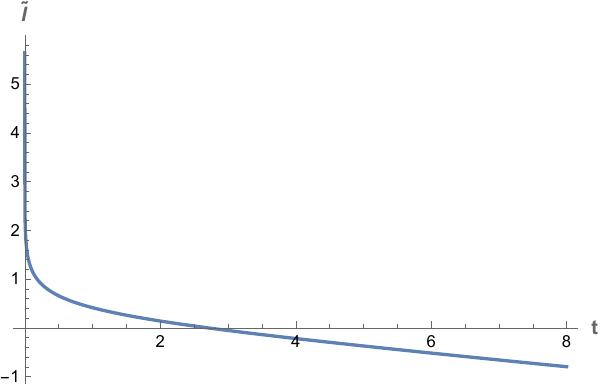}{}\caption{Left panel: Phase transition from $C(2,2)\to D(4,1)$ at $t_c=6.94$ for $T_1=-99.01$. Right panel: Phase transition from $C(3,3)\to D(4,1)$ occurring at $t_c=2.74$ for $T_1=-98.01$. }
    \label{fig:MI4MI5}
\end{figure}
However, there are two connected configurations that arise in the above-mentioned region of validity $ -(\tau+t)<T_1<-\tau$, i.e. for the same disconnected phase $D$(4,1). All of these configurations have been illustrated in figure \ref{fig:4d4c5c}. The first of them pertains to the $C$(2,2) connected configuration, characterized by the fact that the (common) mid-point of both the transitioning intervals $(T_4,T_1) $ and $(T_3,T_2)$ lies below the null shell.
    Therefore, the region of validity of this connected phase splits the region of validity of disconnected phase into two equal time intervals, the first one being
    \begin{align}
        -(\tau+t)<T_1<-\left(\tau+\frac{t}{2}\right)\,. \nonumber
    \end{align}
The  TEE for this $C$(2,2) connected configuration is given by \eqref{C} with $(n,m)=(2,2)$
      \begin{align}
        \tilde{S}_{4C}
&=\frac{c}{6}\log \left(\frac{\beta }{\pi \epsilon }\sinh \left(\frac{\pi}{\beta}\left( 2\tau+t\right)\right)\right)+\frac{c}{6}\log \left(\frac{\beta }{\pi \epsilon }\sinh \left(\frac{\pi}{\beta}t\right)\right)+\frac{c}{6}\log \left(\frac{ 2\tau+t}{\epsilon}\right)\nonumber\\&\hspace{1cm}+\frac{c}{6}\log \left(\frac{ t}{\epsilon}\right)+\frac{c}{6}
\Phi_2(T_1,T_4,\beta)+\frac{c}{6}
\Phi_2(T_2,T_3,\beta)+\frac{c}{3}\pi i\,.
    \end{align}

On the other hand, the range of validity of the other configuration in which  $D(4,1)$ phase comes with $C(3,3)$ is $$-\left(\tau+\frac{t}{2}\right)<T_1<-\tau\,.$$ This happens when the common mid-point of the transitioning intervals $(T_4,T_1) $ and $(T_3,T_2)$ lies above the null shell. The total TEE for $C$(3,3) connected phase is given by
 \begin{align}
        \tilde{S}_{5C}
     &=\frac{c}{6}\log \left(\left(\frac{\beta }{\pi \epsilon }\right)\sinh \left(\frac{\pi}{\beta}\left(  2\tau++t\right)\right)\right)+ \frac{c}{6}\log \left(\left(\frac{\beta }{\pi \epsilon }\right)\sinh \left(\frac{\pi}{\beta} t\right)\right)+\frac{c}{6}\log \left(\frac{  2\tau+ t}{\epsilon}\right)\nonumber\\&\hspace{1cm}+\frac{c}{6}\log \left(\frac{  t}{\epsilon}\right)+\frac{c}{2}\pi i\,.
    \end{align}
  
Once again, there are critical separation values ($t_c$) in which both the connected phases $C(2,2)$ and $C(3,3)$ transitions to the same disconnected phase $D$(4,1). The respective phase transitions are shown in the left and right panels of figure \ref{fig:MI4MI5}.

\subsection{Fifth configuration ($D(4,2) - C(3,4)$)}

\begin{figure}[H]
    \centering
    \includegraphics[width=0.3\linewidth]{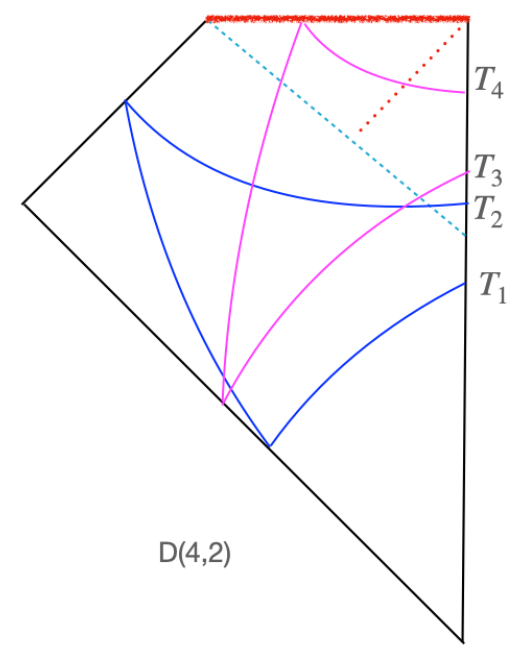}\hspace{3cm}
    \includegraphics[width=0.3\linewidth]{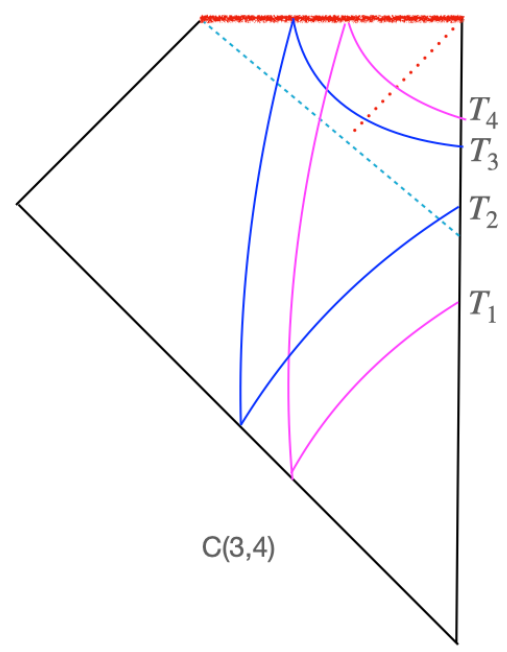}
    \caption{$D(4,2)-C(3,4)$ geodesic configurations valid through $ -\tau<T_1< -\frac{\tau}{2}$.}
    \label{fig:6d6c}
\end{figure}

As the subregion $B$ evolves further and fully transitions to the case 4, it stays there for the rest of the evolution. Therefore, the next variety of cases are only governed by the evolution of the subregion $A$ alone.
Continuing from the previous case, as the subregion $A$ moves upwards, its upper endpoint $T_2$ becomes positive. Initially however $T_2<|T_1|$, thus the mid-point of the transitioning interval $(T_2,T_1)$ is still below the null shell. The corresponding disconnected phase is therefore $D$(4,2) (left panel of figure \ref{fig:6d6c}), and it is valid through
\begin{align}
   -\tau<T_1< -\frac{\tau}{2}\,. \nonumber
\end{align}

\begin{figure}[H]
    \centering
    \includegraphics[width=0.5\linewidth]{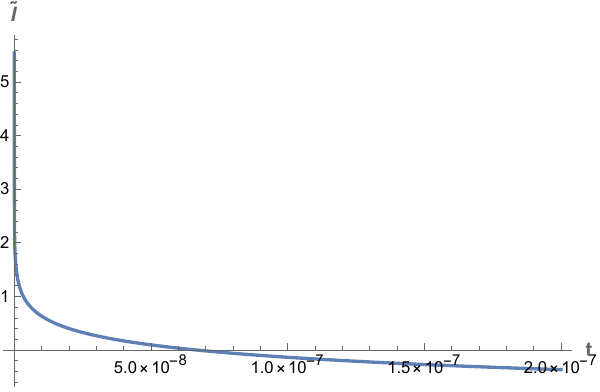}
    \caption{The phase transition from $C(3,4) \to D(4,2)$ occurs immediately at $t_c=6.5\times10^{-8}$ for $T_1=-99.01$.}
    \label{fig:p6}
\end{figure}

In this case, the total TEE is given by \eqref{D} for $(n,m)=(4,2)$ 
  \begin{align}
         \tilde{S}_{6D}
&= \frac{c}{2} \log \left(\frac{\beta }{\pi \epsilon }\sinh \left(\frac{\pi}{\beta} \tau\right)\right)+\frac{c}{6}\log\left(\frac{ \tau}{ \epsilon}\right)+\frac{c}{6}\Phi_4(T_3,T_4,\beta)+\frac{c}{6}
\Phi_2(T_1,T_2,\beta)+\frac{5c}{12}\pi i\,.
    \end{align}
The corresponding connected configuration is the one in which the outer subregion ($T_4,T_1$) is in case 3, while the smaller subregion ($T_3,T_2$) is in case 4. We therefore label this connected configuration by $C(3,4)$ (the corresponding setup appears on the right panel of figure \ref{fig:6d6c}). 
The total timelike entanglement entropy for the connected phase is therefore given by plugging $(n,m)=(3,4)$ in \eqref{C}
    \begin{align}
         \tilde{S}_{6C}
           &=\frac{c}{6} \log \left(\frac{\beta }{\pi \epsilon }\sinh \left(\frac{\pi}{\beta}\left( 2\tau+t\right)\right)\right) +\frac{c}{3} \log \left(\frac{\beta }{\pi \epsilon }\sinh \left(\frac{\pi}{\beta}t\right)\right)+\frac{c}{6}\log \left(\frac{  2\tau+t}{\epsilon}\right) \nonumber\\&\hspace{1cm}+\frac{c}{6}\Phi_4(T_2,T_3,\beta)+\frac{c\pi i}{2}\,.
    \end{align}
      The evolution of the resulting timelike mutual information has been shown in the adjoining figure \ref{fig:p6}. 
It should be noticed that this configuration is a very sensitive one in which the phase transition happens immediately at an extremely small separation. 
 
\subsection{Sixth configuration ($D(4,3) - C(3,4)$)}

\begin{figure}[H]
        \centering
        \includegraphics[width=0.3\linewidth]{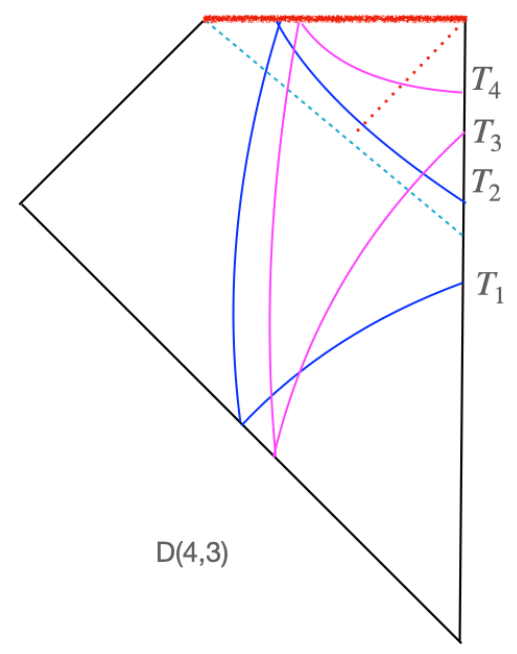}
        \hspace{3cm}
        \includegraphics[width=0.3\linewidth]{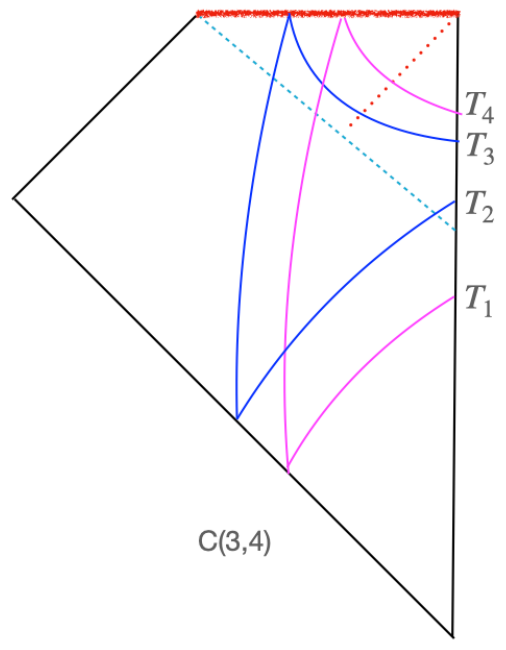}
        \caption{$D(4,3)-C(3,4)$ geodesic configurations valid through $ -\frac{\tau}{2}<T_1<0$.}
        \label{fig:7d7c}
    \end{figure}

In the next disconnected phase, $T_1$ is still below the null shell while $T_2>|T_1|$. This resulting configuration is therefore $D$(4,3) (see left panel of figure \ref{fig:7d7c}), and the temporal range of validity of this configuration is 
\begin{align}
    -\frac{\tau}{2}<T_1<0\,. 
\end{align}
    
The total timelike entanglement entropy of this disconnected phase can be worked out from \eqref{D} for $(n,m)=(4,3)$ to be
    \begin{align}
        \tilde{S}_{6D}
        &= \frac{c}{2} \log \left(\frac{\beta }{\pi \epsilon }\sinh \left(\frac{\pi}{\beta}\tau\right)\right)+\frac{c}{6}\log \left(\frac{ \tau }{\epsilon}\right)\nonumber +\frac{c}{6}\Phi_4(T_3,T_4,\beta)+\frac{c}{2}i\pi\,.
    \end{align}
    \begin{figure}[H]
        \centering
        \includegraphics[width=0.5\linewidth]{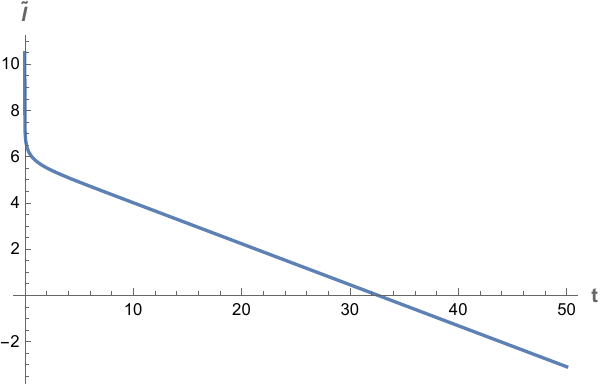}
        \caption{Phase transition from $C(3,4)\to D(4,3)$ occurring at $t_c=32.6$ for $T_1=-49$.
        }
        \label{fig:MI7}
    \end{figure}
Whereas, the corresponding connected phase does not switch and retains the former configuration $C$(3,4) (in other words, $\tilde{S}_{7C}=\tilde{S}_{6C}$; see the right panel of figure \ref{fig:7d7c}) for the entire duration of 
    \begin{align*}
    -\tau<T_1<0\,. \label{6Crange}
\end{align*}
The resulting plot for TMI and the critical time $t_c$ is given in figure \ref{fig:MI7}.

\subsection{Seventh configuration ($D(4,4) - C(4,4)$)} 

This final configuration manifests in the late time limit, when both the subsystems are in the case 4. This configuration is uniquely characterized by $T_1>0$, and the total TEE for $D$(4,4) configuration is  
\begin{align}
        \tilde{S}_{8D}
        &= \frac{2c}{3} \log \left(\frac{\beta }{\pi \epsilon }\sinh \left(\frac{\pi}{\beta}\tau\right)\right) + \frac{c}{3} \log \left(\frac{\beta }{\pi \epsilon }\sinh \left(\frac{\pi}{\beta} \tau\right)\right) +\frac{c}{6}\Phi_4(T_3,T_4,\beta)+\frac{c}{6}\Phi_4(T_1,T_2,\beta)+\frac{c\pi i}{2}\,. \end{align}
        \begin{figure}[H]
            \centering
            \includegraphics[width=0.3\linewidth]{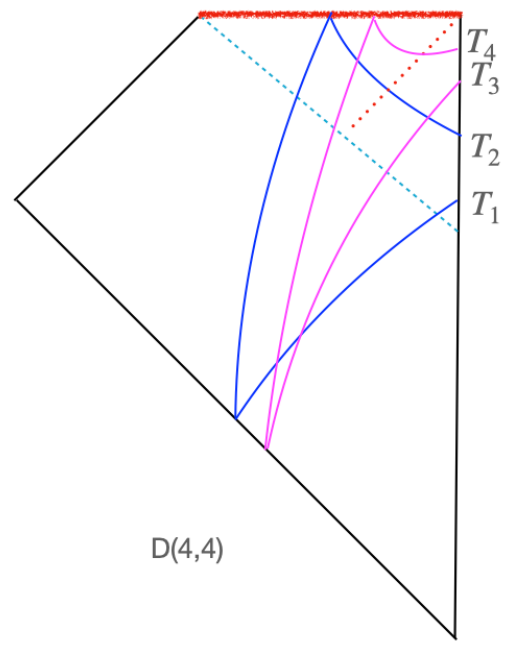}
            \hspace{3cm}
            \includegraphics[width=0.3\linewidth]{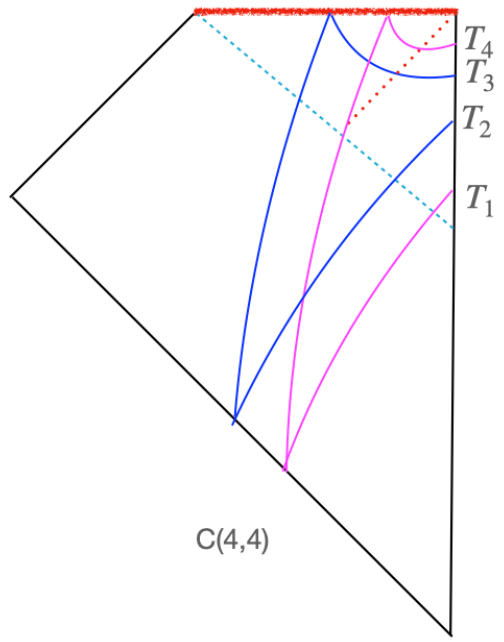}
            \caption{$D(4,4)-C(4,4)$ geodesic configurations valid through $T_1>0$.}
            \label{fig:8d8c}
        \end{figure}
The corresponding connected phase is also characterized by $T_1>0$ and pertains to $C$(4,4) configuration  with 

   \begin{align}
        \tilde{S}_{8C}
       &= \frac{c}{3} \log \left(\frac{\beta }{\pi \epsilon }\sinh \left(\frac{\pi}{\beta}\left( 2\tau+t\right)\right)\right)+ \frac{c}{3} \log \left(\frac{\beta }{\pi \epsilon }\sinh \left(\frac{\pi}{\beta} t\right)\right) +\frac{c}{6}\Phi_4(T_1,T_4,\beta)\nonumber\\&\hspace{1.5cm} +\frac{c}{6}\Phi_4(T_2,T_3,\beta)+\frac{c\pi i}{2}\,.
        \end{align}  
        \begin{figure}[H]
    \centering
    \includegraphics[width=0.5\linewidth]{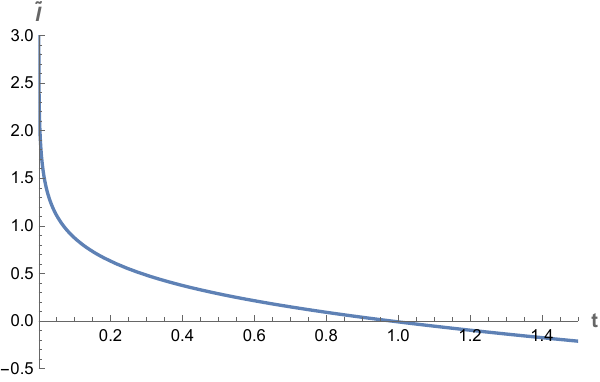}
    \caption{The phase transition from $C(4,4)\rightarrow D(4,2)$ occurring at $t_c=0.98$ for $T_1=0.001$.}
    \label{fig:MI8}
\end{figure}
Once again, figure \ref{fig:8d8c} gives a schematic arrangement of all the surfaces in the Penrose diagram and figure \ref{fig:MI8} depicts the corresponding transition of TMI at the critical point $t_c$.

\subsection{Evolution of timelike mutual information for the Vaidya spacetime }
 \begin{figure}[H]
     \centering
     \includegraphics[width=0.6\linewidth]{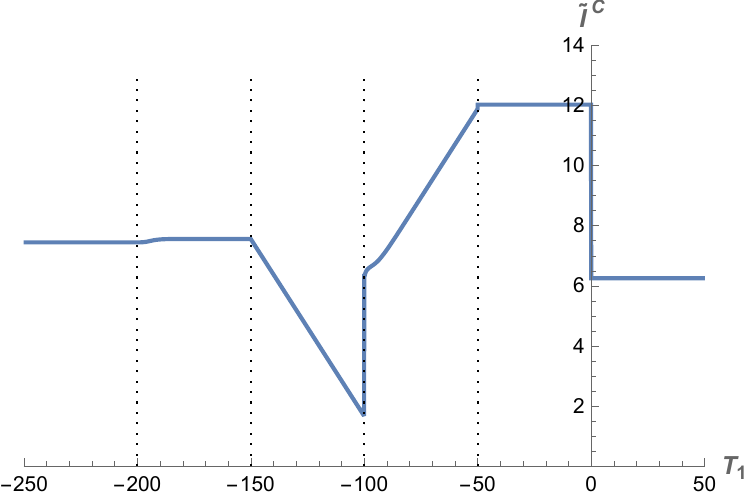}
     \caption{Time evolution of $\tilde{I}^C\geq0$ in the connected phase when subsystem separation is chosen to be $t=10^{-8}$. Vertical partitions are same as the ones in figure \ref{fig:MItimeline}.}
     \label{fig:TMIevolution}
 \end{figure}
In the above section, we have listed out all the possible configurations and arrangements for the pair of timelike intervals, and we have shown that in all of those configurations, the timelike mutual information behaves as expected. Namely, even for the dynamical case, as the timelike intervals get closer, there is a universal phase transition from the disconnected to the connected phase. 
Here, we study a slightly different behaviour in which we will permanently anchor the intervals $A$ and $B$ near enough to be in the connected phase for all configurations, and evolve the combined interval-pairs with respect to time. Our main result here is the plot in figure \ref{fig:TMIevolution}. 
 Our examination of TMI reveals that its general characteristics are at par with the spacelike mutual information, being positive and non-monotonous function of time \cite{Allais:2011ys}. Moreover, just like the spacelike case, its behaviour also depends on the configuration of two disjoint regions; as the subregions are moved farther TMI eventually becomes zero. 
 \\
 
\begin{figure}[H]
     \centering
     \includegraphics[width=0.6\linewidth]{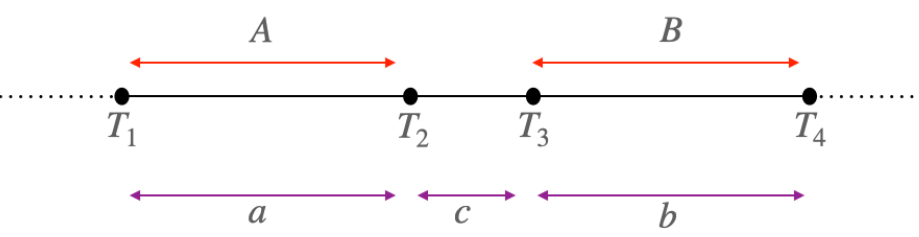}
     \caption{The fact that the connected phase has surfaces homologous to $(T_1,T_4)$ and $(T_2,T_3)$ (and not $(T_1,T_3)$ and $(T_2,T_4)$) implies weak monotonicity.}
     \label{fig:WM}
 \end{figure}

Finally, we have also considered the possibility of an alternative connected phase in which the geodesics are instead homologous to subregions $(T_1,T_3)$ and $(T_2,T_4)$ (for spacelike case, such an investigation appears in \cite{Balasubramanian:2011at}). The corresponding connected phases $C(n,m)$ now go through a slightly different set of orderings. A straight-forward and similar inspection shows that the alternative phase can never be the dominant phase. In other words 
\begin{equation}
    \tilde{S}(A)+\tilde{S}(B)\leq \tilde{S}(A\cup c)+\tilde{S}(B\cup c)\,,
\end{equation}
where in the setup of our non-overlapping intervals $A$ and $B$, we have called the in-between interval as $c$ (see figure \ref{fig:WM}). Renaming $A=a$ and $B=b$, we see that this is nothing but the statement of weak monotonicity condition for three intervals $a,b$ and $c$.
\begin{equation}
    \tilde{S}(a)+\tilde{S}(b)\leq \tilde{S}(A\cup c)+\tilde{S}(B\cup c)\,.
\end{equation}
\section{Probing Subadditivity and Strong Subadditivity using TEE}\label{sec: SSA}

For this section, we now return to the questions of the validity of subadditivity and strong subadditivity (SSA) for two identical timelike subregions in dynamical holography, especially when the subregions overlap. As  mentioned in section \ref{sec:intro}, the standard understanding seems to be that even though the subadditivity (and the statement of Araki-Lieb inequality) is satisfied for the timelike case (and for pseudo entropy), the statement of SSA is violated in general. This was reflected in e.g. \cite{Jiang:2025pen,Afrasiar:2025eam}, although to the best of our knowledge there is only one worked out example of this fact, appearing in \cite{Mollabashi:2021xsd}. However, as we will see in this section, our current approach will furnish more general setup and example where this seems to be true.
\\

We will probe these statements by going beyond what we considered in the previous sections. Whereas in section \ref{sec: TMI} we considered intervals $A$ and $B$ to be strictly disjoint, we now allow for their overlap. In particular, we now have to allow for all possible overlaps that one can have -  from the null set $A\cap B=\Phi$ to the entirety of the intervals, $A\cap B=\tau$.\footnote{Once again, in order to simplify our life, we have only considered equal-sized subregions, as otherwise we will be presented with even more distinct combinations of configurations. As before, the physics of these additional configurations is already contained within the cases we will deal with here.} In fact, this time we will list out the resulting configurations by looking at the amount of overlap we have between the two subsystems. Calling the overlap $t=T_2-T_3\geq0$, we will divide our analysis in two distinct categories: once for $t<\frac{\tau}{2}$ and the other for $t>\frac{\tau}{2}$.
\\

Our treatment of SSA for timelike intervals will involve only the real part of TEE in AdS$_3$-Vaidya. In other words, just like we did in the case of timelike mutual information, we will strip away the non-dynamical (constant) imaginary part and adopt the statement of SSA to only apply to the real part of TEE in the following form:\footnote{For three intervals $a,b$ and $c$, the SSA statement has the usual form $\tilde{S}(a\cup b)+\tilde{S}(b\cup c)\geq \tilde{S}(b)+\tilde{S}(a\cup b\cup c)$. This is nothing but \eqref{ssa}, when $A=a\cup b$ and $B=b\cup c$, i.e. for our current overlapping case.}
\begin{align}
    \tilde{S}(A)+\tilde{S}(B)\geq \tilde{S}(A\cup B)+\tilde{S}(A\cap B)\,. \label{ssa}
\end{align}
In fact, in the next subsections, we will separately plot the LHS and RHS of this SSA inequality, and investigate whether the SSA is satisfied or not.
Additionally, we will also be interested in verifying a weaker form of entanglement inequality called the Araki-Lieb inequality\footnote{Only the second part of the inequality is the Araki-Lieb inequality (which is valid for overlapping subregions as well), whereas the first part is the statement of subadditivity for disjoint subregions. However, sometimes the entire relation is also known as subadditivity. Note that, if \eqref{ssa} is satisfied, the first part of \eqref{AL} is automatically satisfied. However, the non-trivial statement for the timelike case is that the first part of \eqref{AL} is satisfied, even though \eqref{ssa} can be violated.}

\begin{align}
   \tilde{S}(A)+\tilde{S}(B)\geq \tilde{S}(A\cup B)\geq |\tilde{S}(A)-\tilde{S}(B)|\label{AL}.
\end{align}
These inequalities are arguably obeyed by the pseudo entropy (TEE), and is therefore expected to be valid for timelike entanglement entropy as well \cite{Mollabashi:2021xsd}. 
\\

Given that we are now allowing for the overlapping intervals, it is a good time to straighten our notations once more. We generally assume two partially overlapping  identical timelike subsystems $A$ and $B$ each of size $\tau$ anchored at the boundary CFT$_2$ in such a manner that $T_4>T_2\geq T_3> T_1$. We then also have $T_3=T_2-t=T_1+\tau-t$ and $T_4=T_3+\tau=T_1+2\tau-t$. We notice that the left hand side of \eqref{ssa} essentially captures the combined TEE for the disconnected phase. On the other hand, the right hand side of \eqref{ssa} involves $A\cup B=(T_1,T_4)$ and $A\cap B=(T_3,T_2)$. Therefore, we will again use the notations $D(n,m)$ and $C(n,m)$ for the disconnected and ``connected"  configurations respectively, such that for the disconnected phase subregion $B$ is in case $n$ and the subregion $A$ falls under the case $m$ (once again, the case numbers are what appears in section \ref{sec:Recap}).
We therefore have the collective TEE of the disconnected phase in $D(n,m)$ configuration as
\begin{align}
    \tilde{S}_{D(n,m)}=\tilde{S}(A)+\tilde{S}(B)=\tilde{S}_{n}(T_3,T_4)+\tilde{S}_{m}(T_1,T_2)\label{d2}\,,
\end{align}
which appears on the LHS of \eqref{ssa}. For the RHS of \eqref{ssa}, similar to what we had in section \ref{sec: TMI}, we will denote the case number of the subregion $A\cup B=(T_1,T_4)$ by $n'$, and the one given by $A\cap B=(T_2,T_3)$ as $m'$.\footnote{This slight change of notation is essentially to distinguish the current connected phase with the one that appeared in the previous section \ref{sec: TMI}, where the intervals were non-overlapping.} We will therefore have the total TMI given by 
\begin{align}
    \tilde{S}_{C(n',m')}=\tilde{S}(A\cup B)+\tilde{S}(A\cap B)=\tilde{S}_{n'}(T_1,T_4)+\tilde{S}_{m'}(T_3,T_2)\,.\label{c2}
\end{align}

In order to evaluate these TEE values we will once again use equations \eqref{case1} through \eqref{case4} with appropriate case numbers and the values of $i,j$, as dictated by the time parameters in the argument. A list of relevant $\tilde{S}_{D(n,m)}$ and $\tilde{S}_{C(n',m')}$ appear in appendix \ref{a}.
\\

In the light of our adopted notation in \eqref{d2} and \eqref{c2}, the statement of SSA in \eqref{ssa} appears as 
\begin{align}
    \tilde{S}_{D(n,m)}\geq\tilde{S}_{C(n',m')}\,. \label{SSA1}
\end{align}
In the subsections that follow, our aim will be to probe the validity of \eqref{SSA1} for all ranges of overlap size $t$, and also as the intervals move as a function of the boundary time $T_1.$ \\

\subsection{Probing subadditivity and strong subadditivity for fixed  $t<\tau/2$ \label{ssat}}

\begin{figure}[H]
     \centering
\includegraphics[width=1\linewidth]{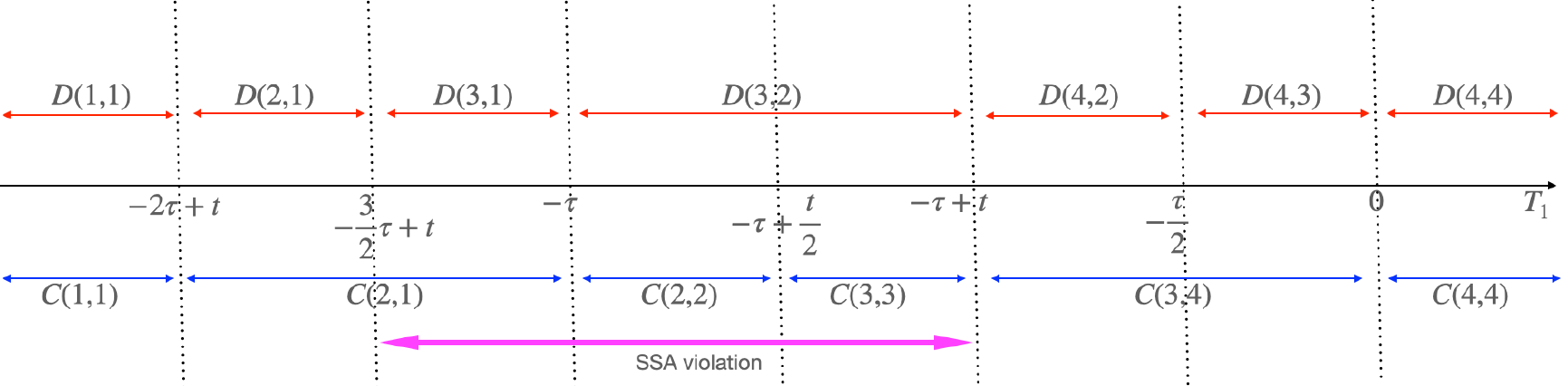}
     \caption{Timeline for small $t$ (not to scale) with respect to $T_1$. In this progression of cases, the magenta line shows the range $\left(-\frac{3}{2}\tau+t<T_1<-\tau+t\right)$ of timelike SSA violation.}
     \label{fig:timeline t}
 \end{figure}

As mentioned in the beginning of section \ref{sec: SSA}, the types of configurations will actually vary as we change the overlap $t$ between the intervals. For this reason, in this subsection, we will focus primarily on the range  $t<\frac{\tau}{2}$ (while $t$ remains fixed somewhere within this range). The relevant zoo of configurations appear in figure \eqref{fig:timeline t}. We will then study the fate of SSA and subadditivity for each of these configurations separately. We will leave the discussion of $t>\frac{\tau}{2}$ to subsection \ref{subsec:SSAlarget}.
\\

 \subsubsection{Early time behaviour for small overlap}
 
\begin{figure}[H]
            \centering
            \includegraphics[width=0.3\linewidth]{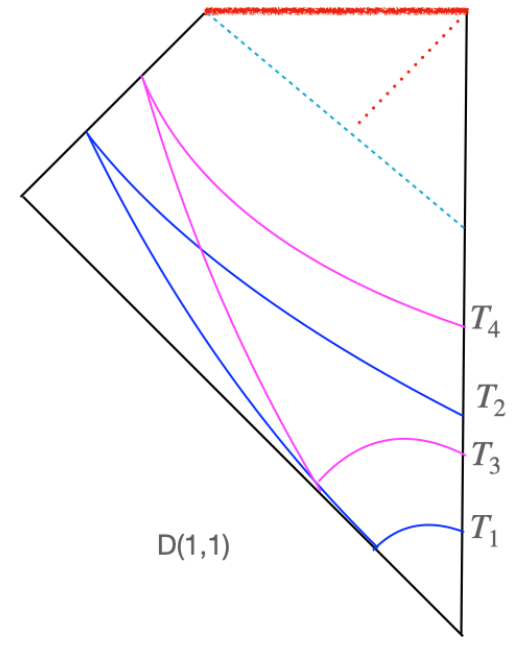}
            \hspace{3cm}
            \includegraphics[width=0.3\linewidth]{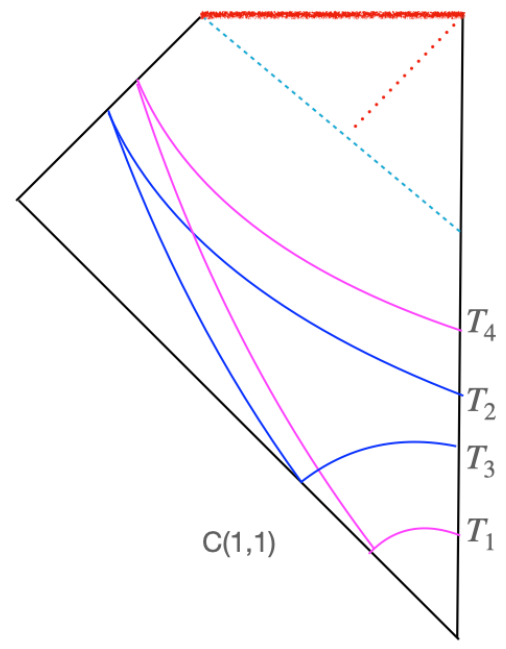}
            \caption{$D(1,1)-C(1,1)$ geodesic configurations valid through $-\infty<T_1<-2 \tau +t$.}
            \label{fig:T1d1c}
        \end{figure}

 In this subsection we will show that the first two configurations for small $t$ obey SSA inequalities over the range $-\infty< T_1<-\frac{3}{2}\tau+t$. The first configuration appearing here is $D(1,1) - C(1,1)$ valid over the boundary time  $-\infty<T_1<-2 \tau +t$  (as shown in \eqref{fig:timeline t}).
 A schematic Penrose diagram for the  $D$(1,1) and  $C$(1,1) configuration is shown in  \eqref{fig:T1d1c}. Because in this case, both the disconnected and the connected phases lie purely within the pure AdS region, we find that the collective TEEs are simply given by \eqref{d11} and \eqref{c11} respectively of the appendix \eqref{a}.

\begin{figure}[H]
            \centering
            \includegraphics[width=0.3\linewidth]{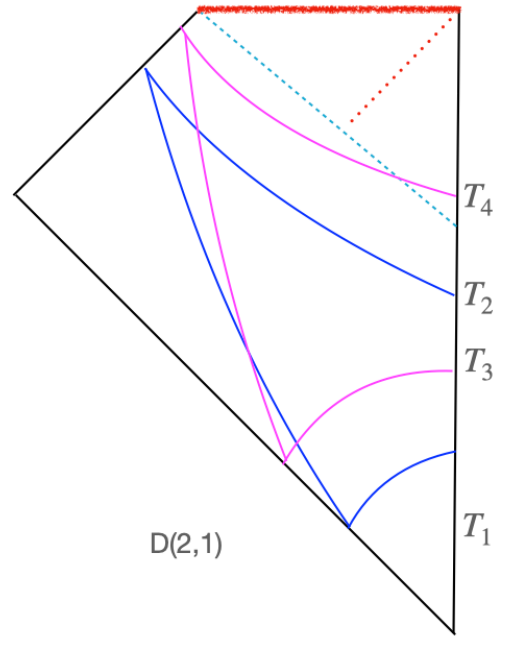}
            \hspace{3cm}
            \includegraphics[width=0.3\linewidth]{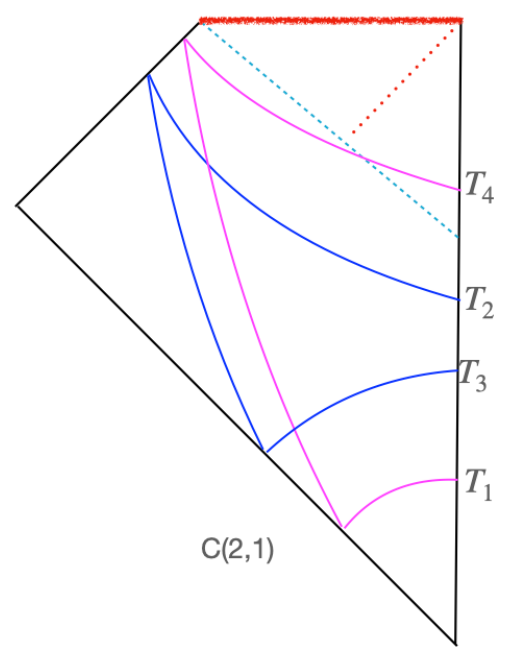}
            \caption{$D(2,1)-C(2,1)$ geodesic configurations valid through $-2 \tau +t<T_1<-\frac{3}{2}\tau+t$.}
            \label{fig:td21}
        \end{figure}

\begin{figure}[htbp]
        \centering
        \includegraphics[width=0.6
        \linewidth]{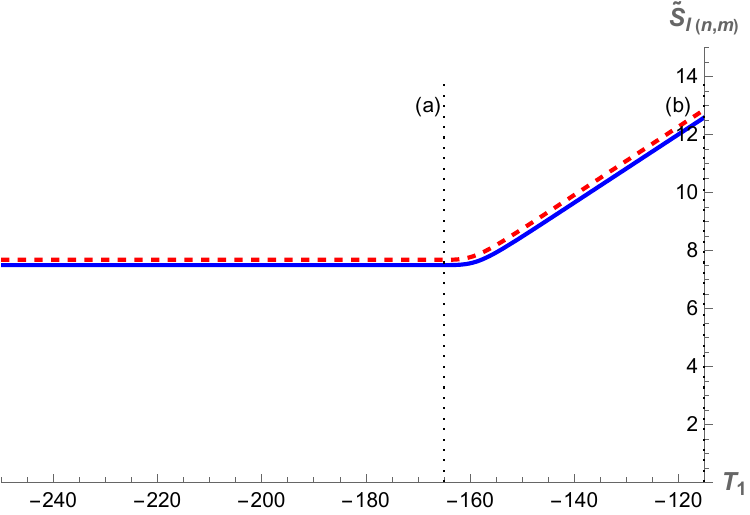}
        \caption{Evolution of boundary intervals in the range $-\infty <T_1<-\frac{3}{2} \tau +t$ for $\tau=100$ and overlap $t=35$. It is manifestly clear that SSA is obeyed throughout in the given range as $\tilde{S}_{C(n',m')}$ (blue curve) is bounded above by $\tilde{S}_{D(n,m)}$ (red dashed curve). In all the similar plots of this section we will label the $y$-axis by $\tilde{S}_{I(n,m)}$ as we have plotted both the connected and the disconnected phases together.}
        \label{fig:ssa1}
\end{figure}

 As the collective region $A\cup B$ is evolved further upwards in time, the disconnected configuration switches over to $D(2,1)$  and the associated configuration representing the RHS of \eqref{ssa} simultaneously switches over to $C(2,1)$. The Penrose diagram of this configuration is depicted in figure \ref{fig:td21}. This combination of configurations lasts till $T_1<-\frac{3}{2}\tau+t$, as given in figure \ref{fig:timeline t}.
The combined plot for these first two configurations have been depicted clearly in figure \ref{fig:ssa1}. The switch-over between the two configurations has been denoted by the vertical dashed line. In this plot (and the ones to follow), the dashed red curve corresponds to $\tilde{S}_{D(n,m)}$ and the blue curve corresponds to $\tilde{S}_{C(n',m')}$. The fact that blue curve is bounded above by the red curve for arbitrarily fixed small $t$, implies that the inequality \eqref{SSA1} is obeyed. Once again while performing the numerical analysis throughout this section, we have fixed common subregion size to be $\tau=100$, mass $M=0.5$ and the UV cutoff to be $\epsilon=10^{-3},$ common to all the configurations. 

 \subsubsection{SSA violating configurations for small $t$}
 
\begin{figure}
    \centering
    \includegraphics[width=0.3\linewidth]{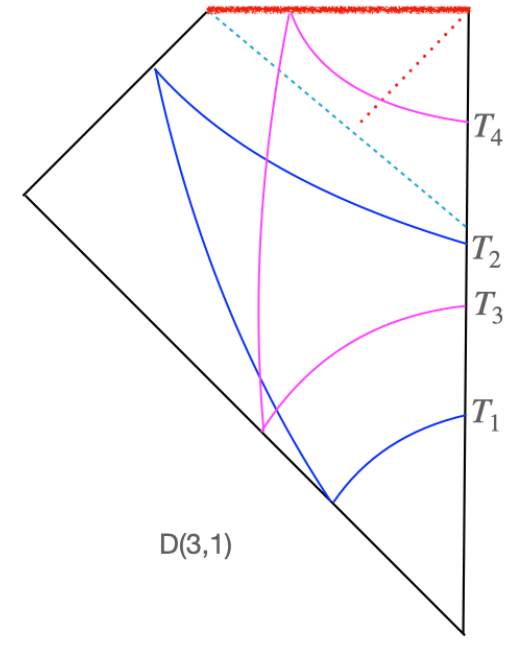} \hspace{3cm} 
     \includegraphics[width=0.3\linewidth]{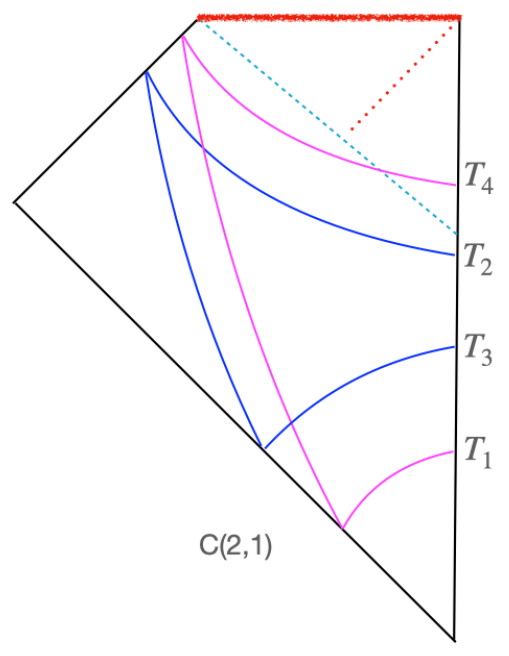}    
    \caption{$D(3,1)-C(2,1)$ geodesic configurations valid through $-\frac{3}{2}\tau+t<T_1<-\tau$.} \label{fig:td3t3c}
    \end{figure}

\begin{figure}
     \centering
     \includegraphics[width=0.3\linewidth]{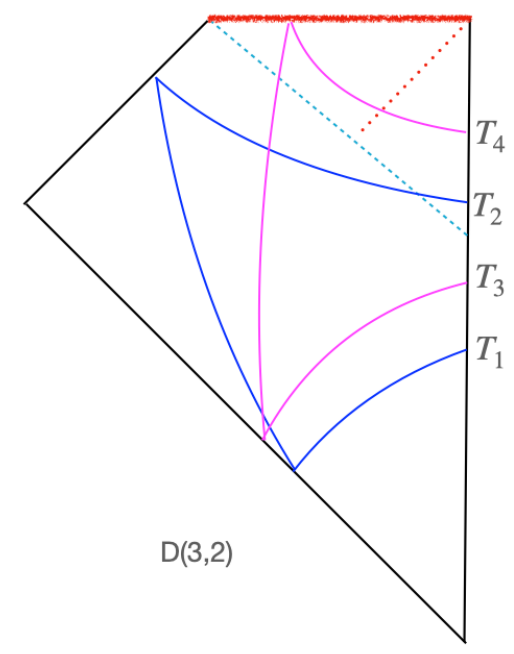}
     \hspace{0.1cm}
     \includegraphics[width=0.3\linewidth]{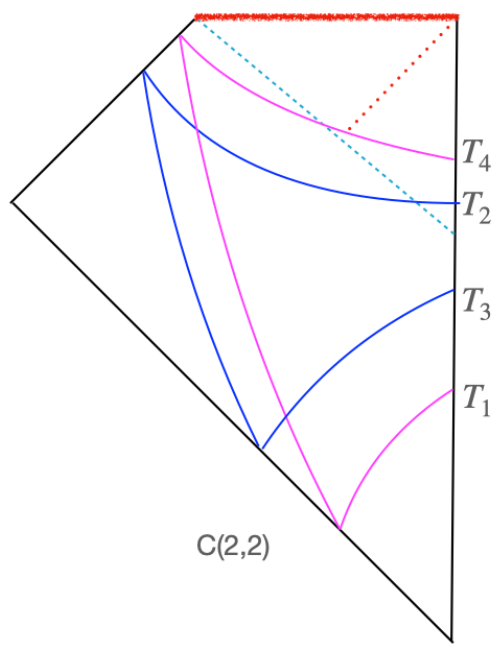}
     \hspace{0.1cm}
     \includegraphics[width=0.3\linewidth]{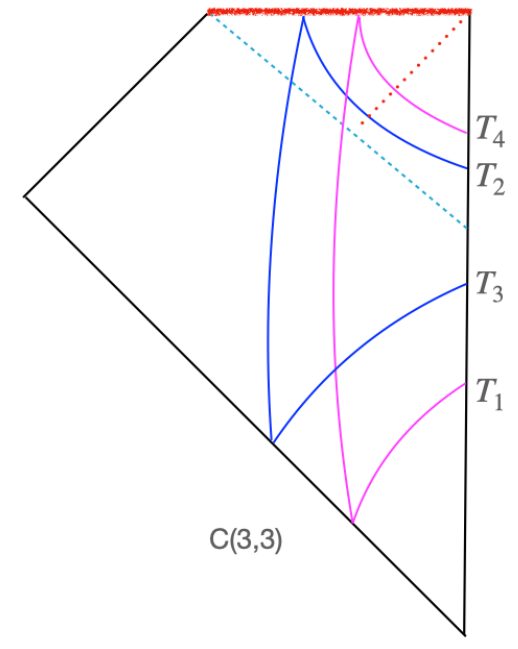}
     \caption{$D(3,2)-C(2,2)$ (left and middle panels), $D(3,2)-C(3,3)$ (left and right panels) geodesic configurations valid through $ -\tau<T_1<-\tau+\frac{t}{2}$ and $-\tau+\frac{t}{2}<T_1<-\tau+t$ respectively.}
     \label{fig:T4d4c5c}
 \end{figure}

 In this subsection, we are going to study the behaviour of SSA in the intermediate boundary times $-\frac{3}{2}\tau+t<T_1<-\tau+t.$
 In the next turn of events, as the subregion $A\cup B$ is further evolved upwards keeping $t$ small, the configuration $D(2,1)$ switches over to $D(3,1)$ the moment $T_1>-\frac{3}{2}\tau+t $ and persists as long as $T_1<-\tau.$ The schematic Penrose diagrams appear in figure \ref{fig:td3t3c}.
As these pairs of subregions are evolved further by increasing $T_1$, the configurations switches from $D(3,1) - C(2,1)$ to $D(3,2) - C(2,2)$, and thereafter to $D(3,2) - C(3,3)$. The relevant Penrose diagrams appear in figure \ref{fig:T4d4c5c}, whereas the relevant time ranges for $T_1$ can be read off from figure \ref{fig:timeline t}. 

\begin{figure}[H]
    \centering
    \includegraphics[width=0.6\linewidth]{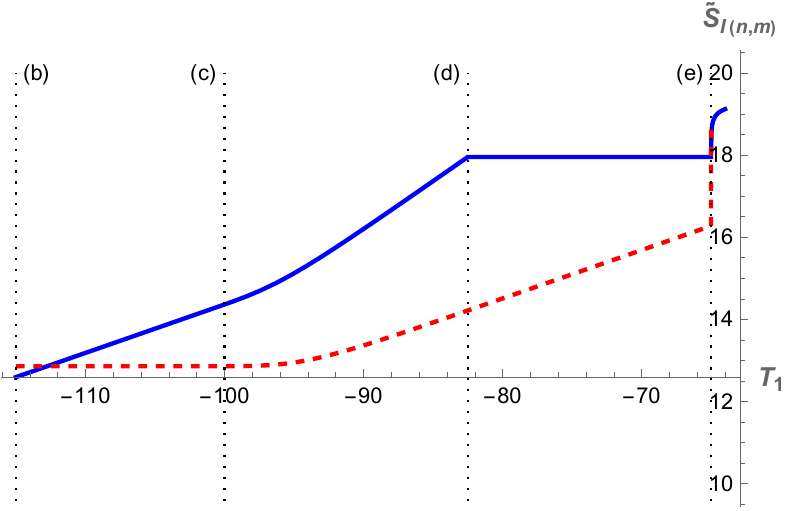}
    \caption{Evolution of SSA during the range $-\frac{3}{2}\tau+t<T_1< -\tau+t$ for $\tau=100$ and overlap $t=35$. Since, the blue curve for $\tilde{S}_{C(n',m')}$ immediately jumps above the red curve representing $\tilde{S}_{D(n,m)}$, SSA is clearly violated in this range.  }
    \label{fig:Inter t}
\end{figure}

The collective TEE of $D(3,2)$ vs $C(2,2)$ and $C(3,3)$ can be computed using equations \eqref{d32}, \eqref{c22} and \eqref{c33} in the appendix \eqref{a}. Once again plotting various $D(n,m)$ and  $C(n',m')$ with the boundary time $T_1$ (see figure \ref{fig:Inter t}), we can clearly clearly see the violation of SSA in the range $-\frac{3}{2}\tau+t<T_1< -\tau+t$. This observation parallels the consensus that TEE being related to the pseudo entropy, generically violates the strong subadditivity \cite{Mollabashi:2020yie,Jiang:2023loq,Afrasiar:2025eam}. 

One may further inquire whether TEE in the same range also reflects a violation in the slightly weaker Araki-Lieb and triangle inequality. It turns out that the answer is negative. In other words, \emph{TEE does obey subadditivity inequality, while simultaneously violating SSA}. We will come to the discussion of subadditivity near the conclusion of this section.

\subsubsection{Late time behaviour for small overlap}

\begin{figure}[H]
          \centering
          \includegraphics[width=0.3\linewidth,height=6cm]{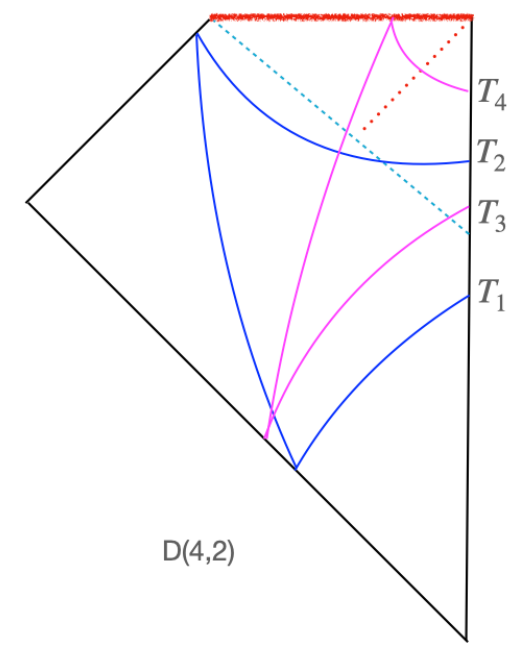}\hfill\includegraphics[width=0.31\linewidth,height=6cm]{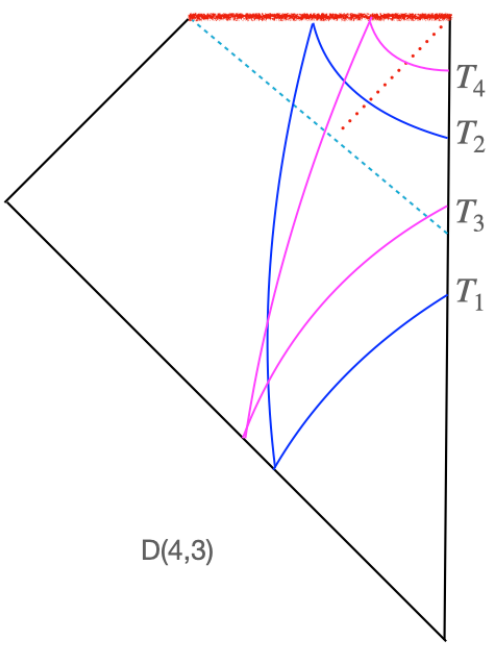}
\includegraphics[width=0.31\linewidth,height=6.2cm]{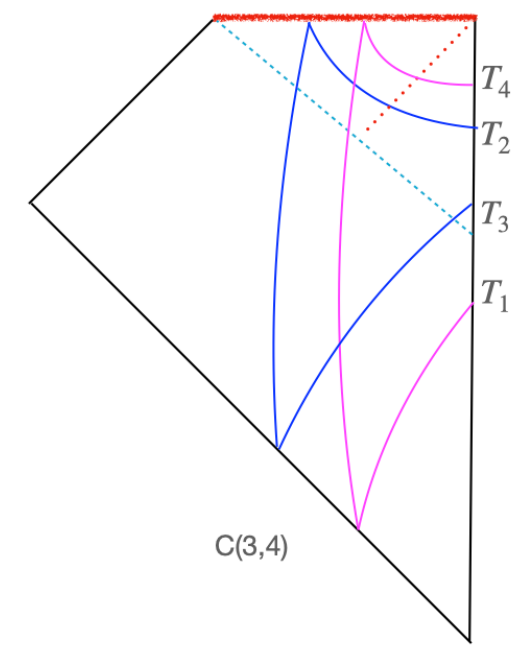}
    \caption{$D(4,2)-C(3,4)$ (left and right panels), $D(4,3)-C(3,4)$ (middle and right panels) geodesic configurations valid through $ -\tau+t<T_1<-\frac{\tau}{2}$ and $-\frac{\tau}{2}<T_1<0$ respectively.}
          \label{fig:t6dT7dt6c}
      \end{figure}

In the later times, from $-\tau+t<T_1<\infty,$  the pair of timelike intervals subsequently goes  through disconnected configurations $D(4,2), \,D(4,3)$ and $D(4,4)$ (for small, fixed overlap $t$) as shown in figure \ref{fig:timeline t}. Initially, the trailing intermediate time configuration $D(3,2)-C(3,3)$ switches over to $D(4,2)-C(3,4)$ at $T_1=-\tau+t$, and that is precisely when the SSA starts being obeyed again. These various configurations are represented in the Penrose diagrams in figure \ref{fig:t6dT7dt6c}.
\\
      
\begin{figure}[H]
    \centering
    \includegraphics[width=0.6\linewidth]{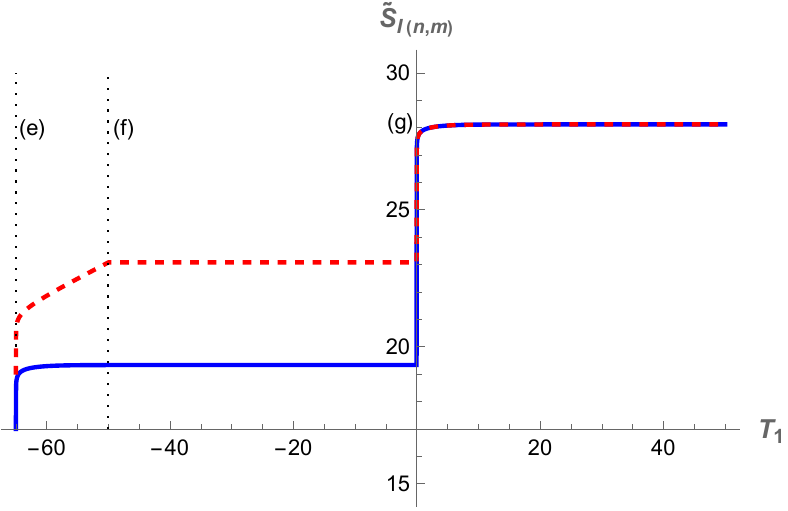}
    \caption{Evolution of boundary intervals in the range $-\tau+t<T_1<0 $ for $\tau=100$ and overlap $t=35$. SSA is restored at $T_1=-\tau+t$ and is obeyed throughout for the future evolution. Even though for positive $T_1$ the blue and red curves seem coincident, the red curve is in fact always above the blue curve.}
    \label{fig:late t}
\end{figure}

Finally, the configuration $D(4,2) - C(3,4)$ switches over to $D(4,4)-C(4,4)$, giving rise to future eternal BTZ geometry. The fact that all these latter configurations obey SSA can be checked from the adjoining combined plot in figure  \ref{fig:late t}.

\begin{figure}[H]
    \centering
    \includegraphics[width=0.6\linewidth]{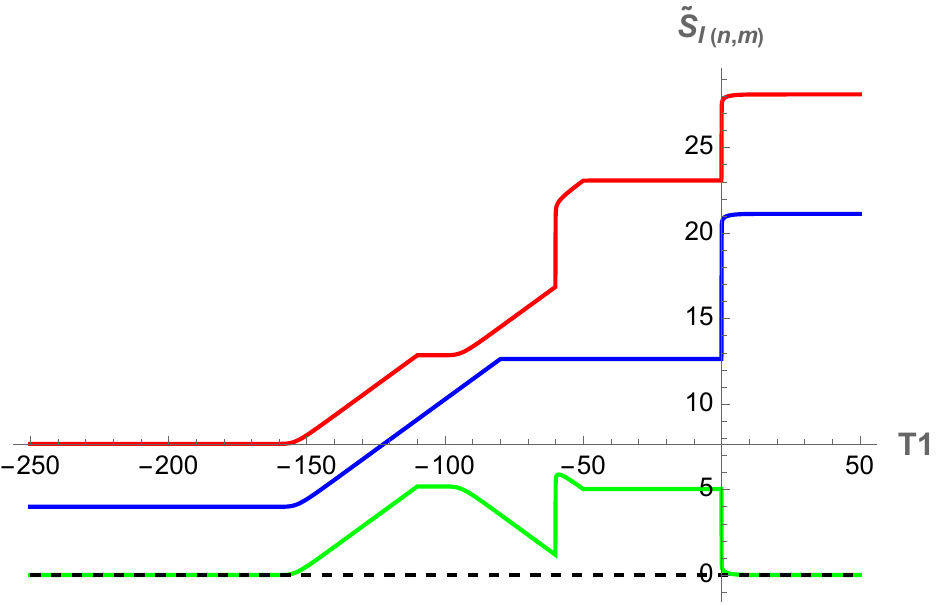}
    \caption{Plots of $\tilde{S}(A)+\tilde{S}(B)$ (in red), $\tilde{S}(A) \cup \tilde{S}(B)$ (in blue) and $|\tilde{S}(A)-\tilde{S}(B)|$ (in green) showing the validity of subadditivity like statement and the Araki-Lieb inequality \eqref{AL} for all $T_1$. We have used $t=40$ and $\tau=100$.}
    \label{fig:tsa}
\end{figure}
 
 Having presented the clear violation of SSA for TEE, one might wonder about the status of the weaker Araki-Lieb and triangle inequality given by \eqref{AL}. This is an important question, as there are no general proofs or arguments behind the violation of the subadditivity, even for the timelike case. In fact, as we mentioned before, the work of \cite{Mollabashi:2021xsd} discusses an example where the subadditivity in the context of pseudo entropy is obeyed. Using our expressions from appendix \ref{a}, we have therefore plotted all parts of \eqref{AL} in figure \ref{fig:tsa}, over all the time ranges given in figure \ref{fig:timeline t}. Here, the red curve represents the usual disconnected phase configuration $\tilde{S}(A)+\tilde{S}(B)$ (LHS of \eqref{AL}), while the blue curve is the middle term of \eqref{AL} (corresponding to the subregion  $A\cup B$) and the green curve represents the far RHS of \eqref{AL} (i.e. $|\tilde{S}(A)-\tilde{S}(B)|$). It is clear from the respective placement of the curves that \eqref{AL} is valid throughout the range of $T_1$.

\subsection{Probing subadditivity and strong subadditivity for fixed $t>\tau/2$}\label{subsec:SSAlarget}

\begin{figure}[H]
     \centering
\includegraphics[width=1\linewidth]{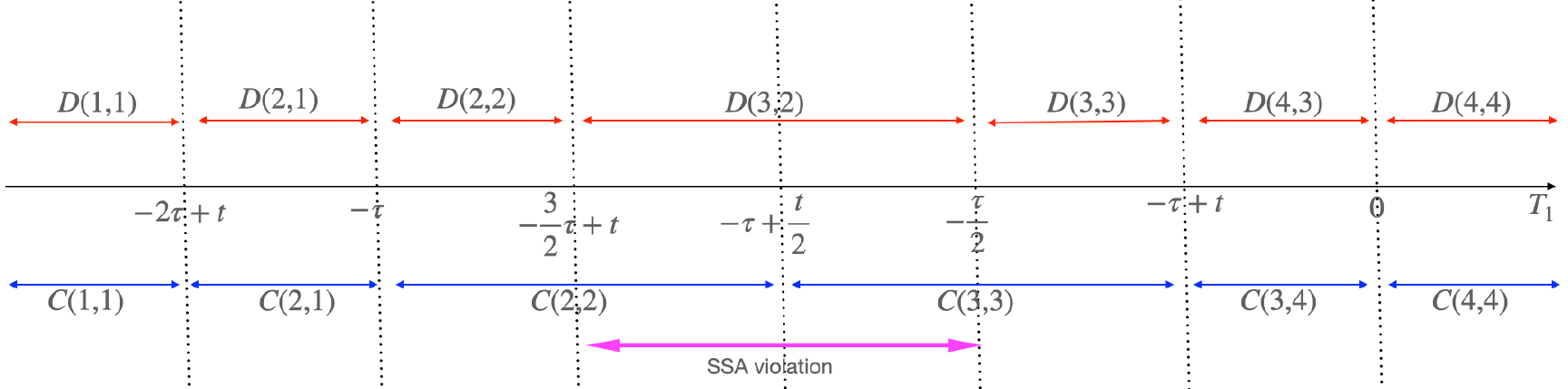}
     \caption{Timeline for large $t$ (not to scale) as function of $T_1$. In this progression of events, the magenta line shows the range  $\left(-\frac{3}{2}\tau+t<T_1<-\frac{\tau}{2}\right)$ of timelike SSA violation.}
     \label{fig:timeline T}
 \end{figure}

\begin{figure}[H]
     \centering
     \includegraphics[width=0.3\linewidth]{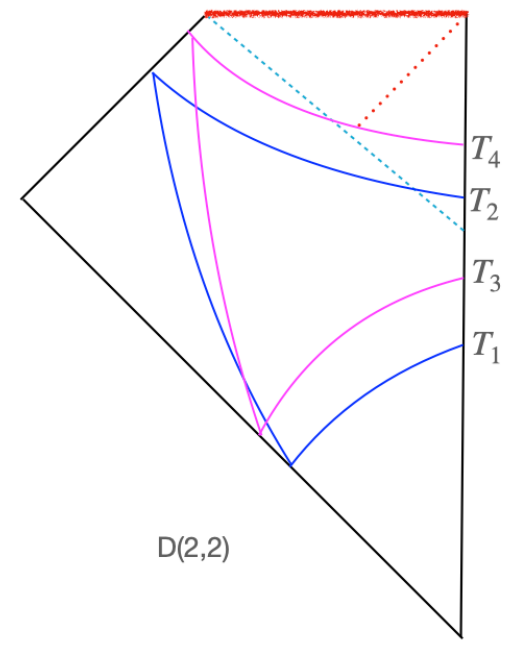}\hspace{2cm}
     \includegraphics[width=0.3\linewidth]{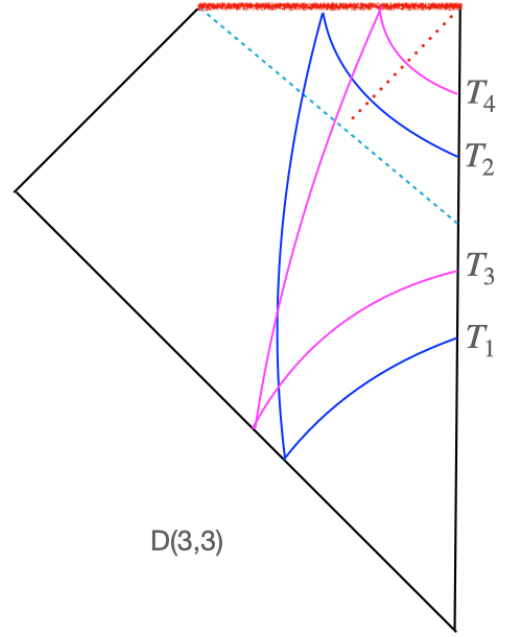}
     \caption{$D$(2,2), $D$(3,3) geodesic configurations valid through $-\tau<T_1<-\frac{3}{2}\tau+t$ and $-\frac{\tau}{2}<T_1<-\tau+t$ respectively.}
     \label{fig:T3dT6d}
 \end{figure}

As advertised before, we now switch our attention to the configurations which appear for $t>\tau/2$. For such large $t$, there again arises seven distinct disconnected configurations. The configurations for the large overlap alter slightly from the small overlap case, as can be seen by comparing figures \ref{fig:timeline t} and \ref{fig:timeline T}. The newest Penrose configurations that appear here have been given in figure \ref{fig:T3dT6d}.
 
As shown in the schematic plot of figure \ref{fig:timeline T}, the region of violation of SSA is now contained within the interval $-\frac{3}{2}\tau+t$ to $-\frac{\tau}{2}.$ Following the simple procedure outlined in the previous section, one can numerically compare the collective TEE of the various configurations, and arrive at the following state of evolution of SSA valid for large overlaps.\\

\begin{figure}[H]
    \centering
    \includegraphics[width=0.6\linewidth]{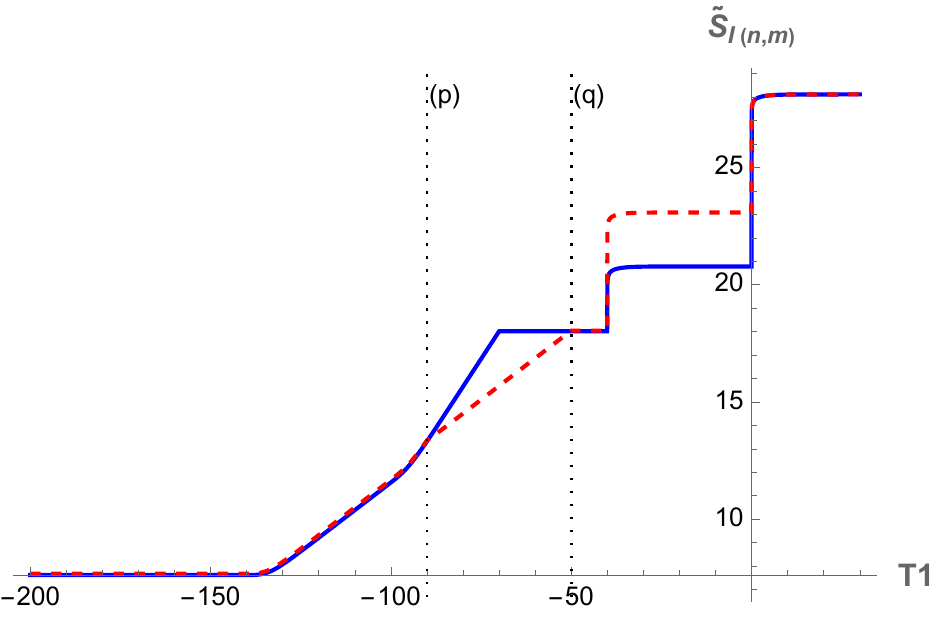}
    \caption{ Evolution of boundary intervals for all range of $T_1$ for large $t=60$ and $\tau=100$. Since, the blue curve for $\tilde{S}_{C(n',m')}$  jumps above the red curve representing $\tilde{S}_{D(n,m)}$, SSA is clearly violated in the range $-\frac{3\tau}{2}+t<T_1<-\frac{\tau}{2}$ shown by (p) and (q) black dotted line. Once again, even if in some places they seem coincident, the red curve is in fact always above the blue curve.}
    \label{fig:SSA}
\end{figure}

We see that during the initial course of evolution from $T_1$ equal to $-\infty$ up to $-\frac{3}{2}\tau+t$, the disconnected  configurations ranges over $D(1,1)-C(1,1)$, $D(2,1)-C(2,1)$ and $D(2,2)-(2,2)$ as shown in the figure \ref{fig:SSA} up to point $(p)$. The collective TEE obeys SSA inequality within this prescribed range. Thereafter, in the intermediate range starting from $(q)=-\frac{3}{2}\tau+t$ until $T_1=-\frac{\tau}{2}, $ the SSA inequality is violated for the configuration $D(3,2)-C(2,2)$ or $C(3,3)$.\footnote{Curiously, comparing figures \ref{fig:timeline t} and \ref{fig:timeline T}, we notice some similarities between the small and large $t$ cases. In both these cases (meaning for all overlaps), the violation of SSA begins at time $T_1=-\frac{3}{2}\tau+t$ and ends when the disconnected phase stops being in the configuration $D(3,2)$.}  For the current scenario of large overlap $t$, the SSA is thereafter restored at point $(q)$ for all the later phases.

Once again, one might ask whether for large $t$ overlaps, the subadditivity or the Araki-Lieb inequality is satisfied or not. We can therefore plot the left, middle and the right hand sides of \eqref{AL} using the suitable equations of appendix \ref{a}. We once again found that for all the configurations mentioned in figure \ref{fig:timeline T}, the triangle inequality is indeed satisfied (see figure \ref{fig:Tsa} below). This provides another set of working examples where the behaviour of SSA and subadditivity are quite different, in accordance with the standard understandings of the timelike case. 
\begin{figure}[H]
     \centering
     \includegraphics[width=0.6\linewidth]{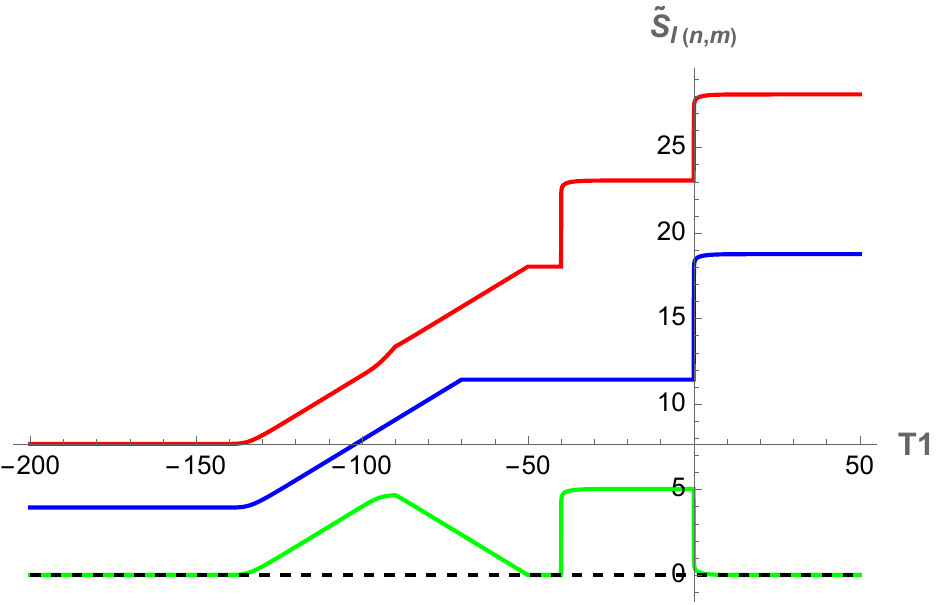}
     \caption{Plots of $\tilde{S}(A)+\tilde{S}(B)$ (in red), $\tilde{S}(A) \cup \tilde{S}(B)$ (in blue) and $|\tilde{S}(A)-\tilde{S}(B)|$ (in green) showing the validity of subadditivity like statement and the Araki-Lieb inequality \eqref{AL} for all $T_1$. We have used $t=60$ and $\tau=100$.}
     \label{fig:Tsa}
 \end{figure}

\subsection{Status of SSA with the variation of overlap}\label{sec:SSAVtchange}
In the previous subsections \ref{ssat} and \ref{subsec:SSAlarget}, we have considered various configurations of the subintervals $A$ and $B$ where the overlap $t$ between them is kept fixed either at a small or a large value, whereas the entire system is dragged up in boundary time. We noticed that for both small $t<\tau/2$ and large $t>\tau/2$, there are certain configurations where the SSA is violated, whereas the weaker triangle inequality is satisfied in every single configurations. In this subsection, we study a slightly different problem, where we vary the overlap $t$ itself, keeping the location of the interval $B$ fixed (in other words, by moving $A$ upwards upon increasing  $T_1$). We will once again see that the SSA violation comes up quite generically in these setups.\footnote{Because in subsections \ref{ssat} and \ref{subsec:SSAlarget}, we saw that subadditivity and the Araki-Lieb inequality are satisfied for all configurations, they are obviously also always obeyed when we vary $t$ in this section.}\\  

\subsubsection{SSA obeying example for all overlap}

\begin{figure}[H]
    \centering
    \includegraphics[width=0.5\linewidth]{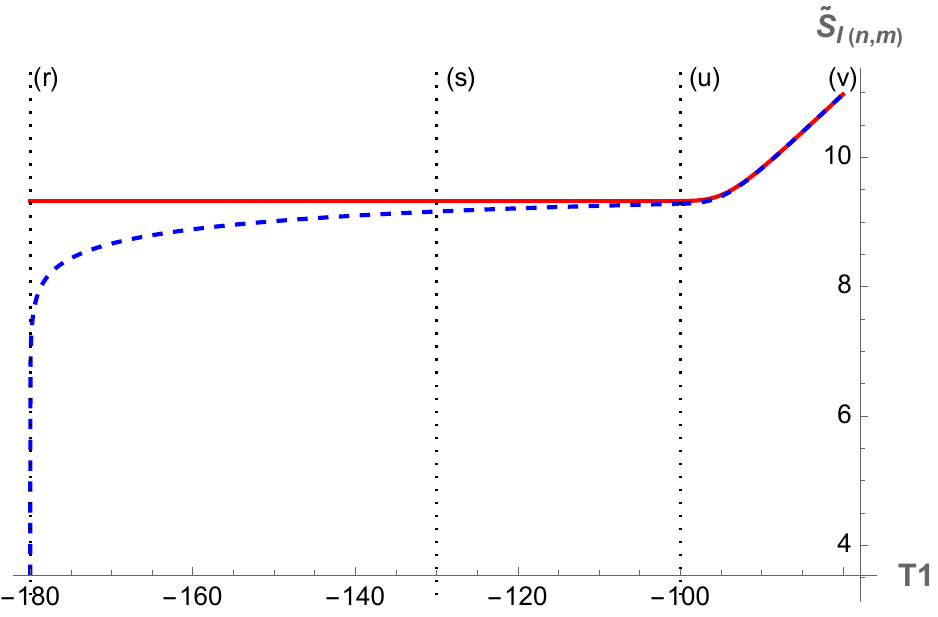}
    \caption{Evolution of boundary interval with varying overlap. The red and the dashed blue curves correspond to the disconnected and connected phases during the evolution.}
    \label{fig:ego}
\end{figure}

Within the scheme mentioned above, we therefore anchor the upper subsystem $B$ at $T_4=20$ and $T_3=T_2=-80$, while the lower mobile subsystem $A$ has the starting lower endpoint of $T_1=-180$. This is a particular numerical choice of $\tau=100$, with a vanishing initial overlap. 
The geometrical  configuration corresponding to this starting situation is $(D(2,1) - C(2,1))$. We now bring the systems closer by gradually sliding $A$ upwards until $T_1\leq-130$ and $T_2\leq-30$. The overlap is now $0\leq t\leq 50$. Throughout this range the SSA is obeyed as can be seen from the curves subtending between points (r) and (s) in figure \ref{fig:ego}. As we dial $A$ further upwards thereby increasing the overlap to more than $50$ until $T_2$ hits $0,$ the persisting configuration is still $D(2,1) - C(2,1)$. However, it now falls under the case of large $t>\tau/2=50$ (the overlap is now over the bigger range of $50\leq  t\leq 80 $). Stitching together the conclusion from the former geodetic configurations,  SSA has now been established for the extended range $0\leq t\leq 80$,  now lying between points (r) and (u) in figure \ref{fig:ego}.\\

In order to check the status on the remaining portion $(80,100),$ we further slide $T_2$ upwards from $0$ to $20$. In doing so, one can notice that the configuration switches from $D(2,1) - C(2,1)$ to $D(2,2) - C(2,2)$ for large $t$. The entire plot and all these above cases have been stitched together and presented in figure \ref{fig:ego}. As it can be clearly seen, for this particular starting configurations, the SSA is obeyed throughout the overlap.
As a special case, in the end we find $\tilde{S}_{D(2,2)}=\tilde{S}_{C(2,2)}$, and the SSA inequality collapses into the  equality because now $t=A\cap B=A\cup B=\tau$. This fact is also immediately visible from the plot of figure \ref{fig:ego} between points (u) and (v), showing the crunching of the blue curve into the red curve from below at the conclusion of the overlapping process. 

\subsubsection{SSA violating example for all overlap} 

\begin{figure}
    \centering
    \includegraphics[width=0.5\linewidth]{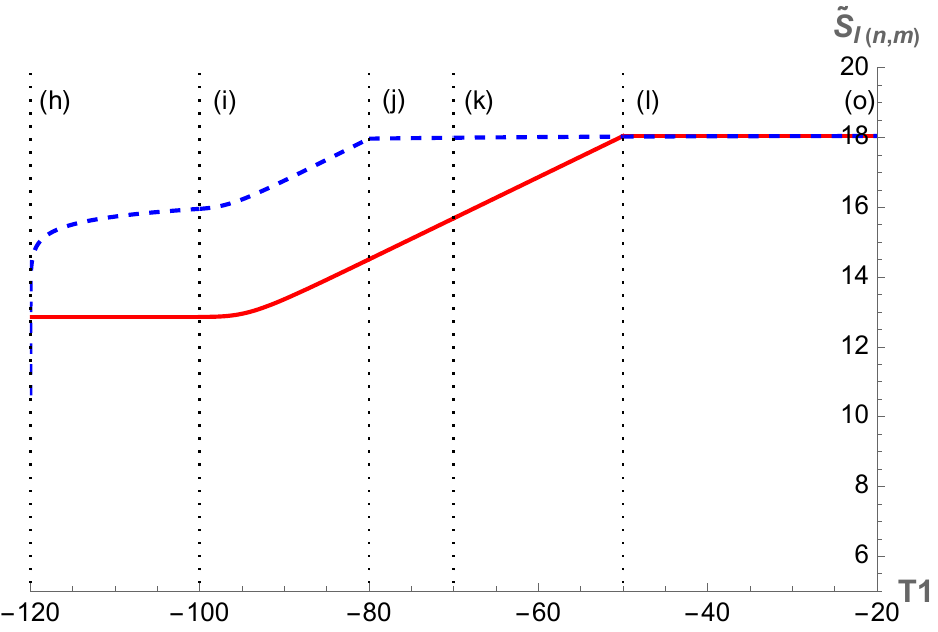}
    \caption{Evolution of boundary intervals with varying overlap showing the violation of SSA between points (h) and (l). The red and the dashed blue curves correspond to the disconnected and connected phases during the evolution.}
    \label{fig:egv}
\end{figure}

In this subsection, we start with a slightly different initial configurations of the intervals $A$  and $B$, where $B$ has been permanently anchored at $T_4=80$ and $T_3=-20$. To start with, this will also be the coordinate of $T_2$ such that $A\cap B=0$, forcing  $T_1=-120$. The geometrical state of this configuration is $D(3,1) - C(2,1)$ for small $t$, and this remains valid during the range $0\leq t\leq 20$ (i.e. until  $T_1\to -100$ and $T_2\to  0$). Upon pushing the interval $A$ even further in $T_1$, the configuration sifts through $D(3,1) - C(2,1)$, $D(3,2)- C(2,2)$ and eventually to $D(3,2)- C(2,2)$. During this progression of events for small $t$ the overlap becomes precisely half of the subsystem size, shown subtended between the points (h) and (k) in the figure \ref{fig:egv}. We see that during the entire contribution the red curve (denoting the left hand side of \eqref{ssa}) is below the blue curve (denoting the right hand side of the SSA equation \eqref{ssa}). Now, the situation pertains to large $t$ values while still being in $D(3,2)-C(3,3)$ configuration lying between (k) and (l).  Finally, the configuration switches to $D(3,3)- C(3,3)$ for which the SSA is once again obeyed between points (l) and (o). 
  
\section{Conclusions}\label{sec:conclude}

The study of pseudo entropy, and especially the related temporal entanglement have been gaining quite some traction in recent years. Not only that it can be studied  from the perspective of AdS/CFT duality  (the starting point considered here) \cite{Doi:2022iyj,Doi:2023zaf}, at an even more fundamental level, there are ongoing investigations stemming from quantum information theory \cite{Milekhin:2025ycm}, and also from Schwinger-Keldysh formalism in the CFT \cite{Gong:2025pnu,Guo:2025ase}. The connection between these various approaches is cleanest within the time  independent states, and it will be extremely important to extend this understandings for various time dependent cases. In our previous and current works, we have tried to provide the holographic entry in this latter direction.\\

To summarize, within the set up of AdS$_3$-Vaidya holography, we have investigated various interplays between quantum extremal surfaces coming from two timelike subregions. Our main probes were quantum information quantities such as mutual information, and various inequalities such as subadditivity and strong subadditivity. Whereas we found that the timelike versions of mutual information and subadditivity continue to show behaviour akin to their spacelike counterparts, the strong subadditivity is usually violated for timelike entanglement. This seems to be the standard lore as discussed in \cite{Mollabashi:2021xsd,Jiang:2025pen,Afrasiar:2025eam} etc. It would be great to clarify why these quantities are treated so differently in such timelike setups. There have also been some recent investigations on the imaginary part of TEE. The imaginary part itself contains crucial information about the commutators of the twist operators, and its first-order temporal derivative \cite{Xu:2024yvf}. Whether the existence of imaginary part is a universal physical feature or an artifact of specific holographic proposals
is still debated in literature \cite{Jiang:2025pen,Zhao:2025zgm,Li:2026fcr}.  It was also claimed to be sensitive to relative phases in transition amplitudes according to a different study \cite{Goki:2026hpl}. For our dynamical case, the imaginary parts could be quite distinct, and their physical significance is not very clear. In fact, curiously enough, we found that SSA is never violated when the imaginary parts mutually cancel out.

\paragraph{Acknowledgements}

$\newline$
\\
The authors would like to thank Haitang Yang for insightful discussion. BS thanks CSIR-HRDG for financial support with fellowship number 09/1022(12312)/2021-EMR-I. The work of GK and DS is supported by SERB sponsored Research Project number CRG/2023/000904.  The work of DS is also supported by the DST-FIST grant number SR/FST/PSI-225/2016.

\appendix

\section{ Collective TEE for various overlapping configurations\label{a}}

This appendix contains some useful formulas that we have utilized in section \ref{sec: SSA}. Essentially they are the analytic expressions for various TEEs that appear in either sides of \eqref{ssa} for various cases, namely in equations \eqref{d2} and \eqref{c2}. 
\begin{align}
\tilde{S}_{D(1,1)}=&\frac{2}{3} c \log \left(\frac{\tau }{\epsilon }\right).\label{d11}
\end{align}

\begin{align}
\tilde{S}_{C(1,1)}=&\frac{1}{3} c \log \left(\frac{2 \tau -t}{\epsilon }\right)+\frac{1}{3} c \log \left(\frac{t}{\epsilon }\right).\label{c11}
\end{align}

\begin{align}
\tilde{S}_{D(2,1)}=&\frac{1}{6} c \log \left(\frac{\beta  \sinh \left(\frac{\pi  \tau }{\beta }\right)}{\pi  \epsilon }\right)+\frac{1}{2} c \log \left(\frac{\tau }{\epsilon }\right)+\frac{c}{6}  \Phi_2(T_3,T_4).\label{d21}\end{align}

\begin{align}
 \tilde{S}_{C(2,1)}=&\frac{1}{6} c \log \left(\frac{\beta  \sinh \left(\frac{\pi  (2 \tau -t)}{\beta }\right)}{\pi  \epsilon }\right)+\frac{1}{6} c \log \left(\frac{2 \tau -t}{\epsilon }\right)+\frac{1}{3} c \log \left(\frac{t}{\epsilon }\right)+\frac{c}{6} \Phi_2(T_1,T_4).\label{c21}\end{align}
\begin{align}
\tilde{S}_{D(3,1)}= &\frac{1}{6} c \log \left(\frac{\beta  \sinh \left(\frac{\pi  \tau }{\beta }\right)}{\pi  \epsilon }\right)+\frac{1}{2} c \log \left(\frac{\tau }{\epsilon }\right).\label{d31}
\end{align}

\begin{align}
\tilde{S}_{D(3,2)}=& \frac{1}{3} c \log \left(\frac{\beta  \sinh \left(\frac{\pi  \tau }{\beta }\right)}{\pi  \epsilon }\right)+\frac{1}{3} c \log \left(\frac{\tau }{\epsilon }\right)+\frac{c}{6}  \Phi_2(T_1,T_2).\label{d32}\end{align}
\begin{align}
\tilde{S}_{C(2,2)}=&  \frac{1}{6} c \log \left(\frac{\beta  \sinh \left(\frac{\pi  (2 \tau -t)}{\beta }\right)}{\pi  \epsilon }\right)+\frac{1}{6} c \log \left(\frac{\beta  \sinh \left(\frac{\pi  t}{\beta }\right)}{\pi  \epsilon }\right)+\frac{1}{6} c \log \left(\frac{2 \tau -t}{\epsilon }\right)+\frac{1}{6} c \log \left(\frac{t}{\epsilon }\right)\nonumber\\&\hspace{0.5cm}+\frac{c}{6}  \Phi_2(T_1,T_4)+\frac{c}{6} \Phi_2(T_3,T_2).\label{c22}\end{align}

\begin{align}
 \tilde{S}_{C(3,3)}=  \frac{1}{6} c \log \left(\frac{\beta  \sinh \left(\frac{\pi  (2 \tau -t)}{\beta }\right)}{\pi  \epsilon }\right)+\frac{1}{6} c \log \left(\frac{\beta  \sinh \left(\frac{\pi  t}{\beta }\right)}{\pi  \epsilon }\right)+\frac{1}{6} c \log \left(\frac{2 \tau -t}{\epsilon }\right)+\frac{1}{6} c \log \left(\frac{t}{\epsilon }\right).\label{c33}\end{align}
\begin{align}
\tilde{S}_{D(4,2)}=    \frac{1}{2} c \log \left(\frac{\beta  \sinh \left(\frac{\pi  \tau }{\beta }\right)}{\pi  \epsilon }\right)+\frac{1}{6} c \log \left(\frac{\tau }{\epsilon }\right)+\frac{c}{6} \Phi_2(T_1,T_2)+\frac{c}{6} \Phi_4(T_3,T_4).\label{d42}\end{align}
\begin{align}
 \tilde{S}_{C(3,4)}=  \frac{1}{6} c \log \left(\frac{\beta  \sinh \left(\frac{\pi  (2 \tau -t)}{\beta }\right)}{\pi  \epsilon }\right)+\frac{1}{3} c \log \left(\frac{\beta  \sinh \left(\frac{\pi  t}{\beta }\right)}{\pi  \epsilon }\right)+\frac{1}{6} c \log \left(\frac{2 \tau -t}{\epsilon }\right)+\frac{c}{6}  \Phi_4(T_3,T_2).\label{c34}
 \end{align}
 \begin{align}
 \tilde{S}_{D(4,3)}=  \frac{1}{2} c \log \left(\frac{\beta  \sinh \left(\frac{\pi  \tau }{\beta }\right)}{\pi  \epsilon }\right)+\frac{1}{6} c \log \left(\frac{\tau }{\epsilon }\right)+\frac{c}{6} \Phi_4(T_3,T_4).\label{d43}\end{align}

\begin{align}
\tilde{S}_{D(4,4)}=  \frac{2}{3} c \log \left(\frac{\beta  \sinh \left(\frac{\pi  \tau }{\beta }\right)}{\pi  \epsilon }\right)+\frac{c}{6} \Phi_4(T_1,T_2)+\frac{c}{6} \Phi_4(T_3,T_4).\label{d44}\end{align}
\begin{align}
\tilde{S}_{C(4,4)}=   \frac{1}{3} c \log \left(\frac{\beta  \sinh \left(\frac{\pi  (2 \tau -t)}{\beta }\right)}{\pi  \epsilon }\right)+\frac{1}{3} c \log \left(\frac{\beta  \sinh \left(\frac{\pi  t}{\beta }\right)}{\pi  \epsilon }\right)+\frac{c}{6} \Phi_4(T_1,T_4)+\frac{c}{6} \Phi_4(T_3,T_2).\label{c44}\end{align}
\begin{align}
\tilde{S}_{D(2,2)}=&\frac{1}{6} c \log \left(\frac{\beta  \sinh \left(\frac{\pi  \left(T_2-T_1\right)}{\beta }\right)}{\pi  \epsilon }\right)
   +\frac{1}{6} c \log \left(\frac{\beta  \sinh \left(\frac{\pi  \left(T_4-T_3\right)}{\beta }\right)}{\pi  \epsilon }\right)+ \frac{1}{6} c \log \left(\frac{T_2-T_1}{\epsilon }\right)\nonumber \\&\hspace{0.5cm}+\frac{1}{6} c \log \left(\frac{T_4-T_3}{\epsilon }\right)
   +\frac{1}{6} c \phi _2\left(T_1,T_2\right)+\frac{1}{6} c \phi _2\left(T_3,T_4\right).\label{d22}\end{align}
\begin{align}
\tilde{S}_{D(3,3)}=&\frac{1}{6} c \log \left(\frac{\beta  \sinh \left(\frac{\pi  \left(T_2-T_1\right)}{\beta }\right)}{\pi  \epsilon }\right)+\frac{1}{6} c \log \left(\frac{\beta  \sinh \left(\frac{\pi  \left(T_4-T_3\right)}{\beta }\right)}{\pi  \epsilon }\right)+ \frac{1}{6} c \log \left(\frac{T_2-T_1}{\epsilon }\right)\nonumber\\&\hspace{0.5cm}+\frac{1}{6} c \log \left(\frac{T_4-T_3}{\epsilon }\right).\label{d33}
\end{align}
\bibliographystyle{utphys}
\bibliography{ref}
\end{document}